\documentclass[12pt,a4paper]{article}
\usepackage{epsfig}
\usepackage{mathtools}
\usepackage{amsfonts}
\usepackage{amssymb}
\usepackage{amsmath}
\usepackage{color}
\usepackage{slashed}
\usepackage{cite}
\usepackage{caption}
\usepackage{subcaption}

\usepackage{geometry} 
\usepackage{graphicx}
\graphicspath{{pictures/}}

\usepackage{multirow}           
\usepackage{array}              

\def\la              {\langle}
\def\ra              {\rangle}

\def\Li2              {\text{Li}_2}

\DeclareMathOperator\arctanh{arctanh}

\begin{document}

\begin{center}

\vspace{3cm}

{\bf \Large Radiative corrections to false vacuum decay in quantum mechanics} \vspace{1cm}

{\large M.A. Bezuglov$^{1,2}$ A.I. Onishchenko$^{1,2,3}$}\vspace{0.5cm}

{\it $^1$Bogoliubov Laboratory of Theoretical Physics, Joint
	Institute for Nuclear Research, Dubna, Russia, \\
	$^2$Moscow Institute of Physics and Technology (State University), Dolgoprudny, Russia, \\
	$^3$Skobeltsyn Institute of Nuclear Physics, Moscow State University, Moscow, Russia}\vspace{1cm}

\abstract{We consider radiative corrections to false vacuum decay within the framework of quantum mechanics for the general potential of the form $\frac{1}{2} M \phi^2 (\phi-A) (\phi-B)$, where $M$, $A$ and $B$ are arbitrary parameters. For this type of potential we provide analytical results for Green function in the background of corresponding bounce solution together with one loop expression for false vacuum decay rate.  Next, we discuss the computation of higher order corrections for false vacuum decay rate and provide numerical expressions for two and three loop contributions. }
\end{center}

\begin{center}
Keywords: false vacuum decays, radiative corrections, quantum mechanics 	
\end{center}

\newpage

\tableofcontents{}\vspace{0.5cm}

\renewcommand{\theequation}{\thesection.\arabic{equation}}

\section{Introduction}

First-order phase transitions driven by scalar fields play an important role in high-energy, astro-particle physics and cosmology. Such first order phase transitions go through the nucleation of new phase bubbles, often around impurities, which subsequently expand. A well known example, which attracted recently a lot of attention, is the possible metastability\footnote{For tunneling rates calculations in Standard Model and its extensions see
\cite{SMtunneling1,SMtunneling2,SMtunneling3,SMtunneling4,SMtunneling5,SMtunneling6,SMtunneling7,SMtunneling8,SMtunneling9} and references therein.} of electroweak vacuum at a scale around $10^{11}$ GeV \cite{SMmetastability1,SMmetastability2,SMmetastability3,SMmetastability4,SMmetastability5,SMmetastability6,SMmetastability7,SMmetastability8}. Next, first-order phase transitions have a potential of producing stochastic gravitational wave backgrounds \cite{StochasticBackground1,StochasticBackground2,StochasticBackground3,StochasticBackground4,StochasticBackground5}, which could be further studied experimentally by existing and forthcoming gravitational wave experiments \cite{GravitationalWavePhysics}. In addition, first order electroweak phase transition may satisfy Sakharov's conditions \cite{SakharovConditions} and be responsible for the generation of baryon asymmetry of our Universe, see \cite{BaryonAsymmetry1,BaryonAsymmetry2} and references therein. The amount of produced baryon asymmetry crucially depends on the dynamics of phase transition. It should be noted, that in Standard Model the electroweak phase transition is not actually first-order, but a crossover \cite{crossover1,crossover2,crossover3,crossover4}. However, electroweak baryogenesis could survive in its extensions, many of which include extra dynamical scalar fields. The role of impurities catalyzing the mentioned first-order phase transitions may be played by small evaporating black holes or other gravitational inhomogeneities, see \cite{PTcatalysis1,PTcatalysis2,PTcatalysis3,PTcatalysis4} and references therein. 

The first rigorous description of the mentioned quantum phase transition between different vacuum states appeared in \cite{Coleman1,Coleman2,Coleman3,Kobzarev}. As is known, the false vacuum decay rate may be related to the imaginary part of ground state energy. However, to calculate the latter one first needs to perform an analytical continuation of the potential so that the false vacuum is stable. This original method was further developed and is known now as {\it potential deformation method}, see for review \cite{KleinertPathIntegrals,IntroQuantumMechanics,QFTCriticalPhenomena,InstantonsLargeN,WeinbergClassicalSolutions,PrecisionDecayRates}. Recently a  more direct way to compute tunneling probabilities using path integrals appeared in \cite{DirectApproachQuantumTunneling}, see also \cite{PrecisionDecayRates}. In both methods the decay rates are given by path integrals around bounce configurations (solutions of the Euclidean equations of motion used to evaluate path integrals in saddle point approximation) divided by corresponding path integrals around static false vacuum (FV) solutions.

At one loop there is a number of different methods to compute functional determinants arising in evaluation of path integrals.  Among them are direct evaluation of spectrum for solvable potentials\footnote{See \cite{InstantonsLargeN} and references therein.}, heat kernel methods \cite{HeatKernelUserManual,CalculationsExternalFields,MassiveContributionsQCDtunneling,DunneFunctionalDeterminants}, Green function methods \cite{KleinertChervyakov1,KleinertChervyakov2,GarbrechtMillington1,GarbrechtMillington2,GarbrechtMillington3} and of course famous Gel'fand-Yaglom method \cite{GenfaldYaglom} and its generalizations \cite{KirstenMcKane1,KirstenMcKane2}. The methods beyond one loop were mostly developed in the study of instantons \cite{ABCinstantons,InstantonsQCD,LecturesInstantons} in quantum mechanics \cite{AleinikovShuryak,Olejnik:1989id,Shuryak1,Shuryak2,Shuryak3}, effective Euler-Heisenberg lagrangian \cite{EulerHeisenberg1,EulerHeisenberg2,EulerHeisenberg3} and corrections to classical string solutions \cite{Tseytlin1,Tseytlin2}.  

The purpose of this paper is to extend the methods of \cite{AleinikovShuryak,Olejnik:1989id,Shuryak1,Shuryak2,Shuryak3} for the computation of higher order radiative corrections to false vacuum decay in quantum mechanics. We organized the paper as follows.  In Section \ref{decayQM} we set up our framework and present bounce solutions for the potential of the most general form containing both cubic and quartic interactions. Next, we derive analytical expression for the Green function in the background of found bounce solution. We were able to determine one-loop false vacuum decay rate analytically using Gel'fand-Yaglom method. Beyond one loop we used Feynman diagram technique on the top of false vacuum and bounce solutions to get two and three loop corrections to decay rate. Finally, in Section \ref{ConclusionSec} we come with our conclusion.

\section{False vacuum decay in quantum mechanics}\label{decayQM}

The perturbative vacuum obtained by small quantum fluctuations around false (metastable) vacuum will eventually decay. This means that the energy of the ground state has a small imaginary part. The most convenient way to compute ground state energy is to consider small temperature behavior of the thermal partition function
\begin{eqnarray}
Z (\beta) = \text{tr}\, e^{-\beta H(\beta)},
\end{eqnarray}
so that
\begin{eqnarray}
E = - \lim_{\beta\to\infty}\frac{1}{\beta} \log Z(\beta) .
\end{eqnarray}
The imaginary part of the energy comes from imaginary part of the thermal partition function\footnote{See for example \cite{InstantonsLargeN}.}
\begin{eqnarray}
\text{Im} E = - \lim_{\beta\to\infty}\frac{1}{\beta}\frac{\text{Im}Z}{\text{Re}Z}.
\end{eqnarray}
\begin{figure}[h]
	\centering
	\begin{subfigure}{.5\textwidth}
		\centering
		\includegraphics[width=0.8\linewidth]{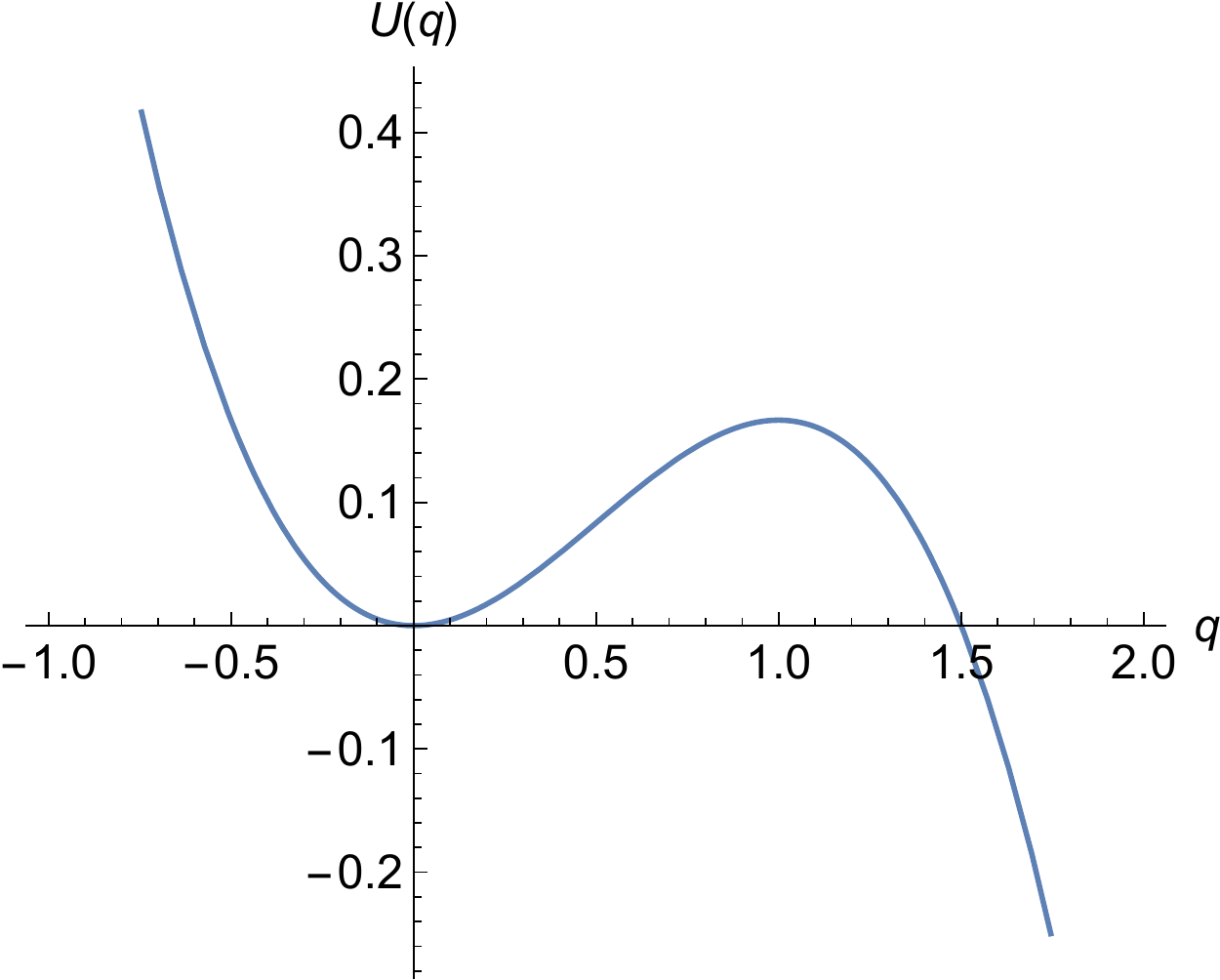}
		\caption{$U(q) = \frac{1}{2}q^2 -\frac{1}{3}q^3$}
		\label{fig:potentialb0}
	\end{subfigure}%
	\begin{subfigure}{.5\textwidth}
		\centering
		\includegraphics[width=0.8\linewidth]{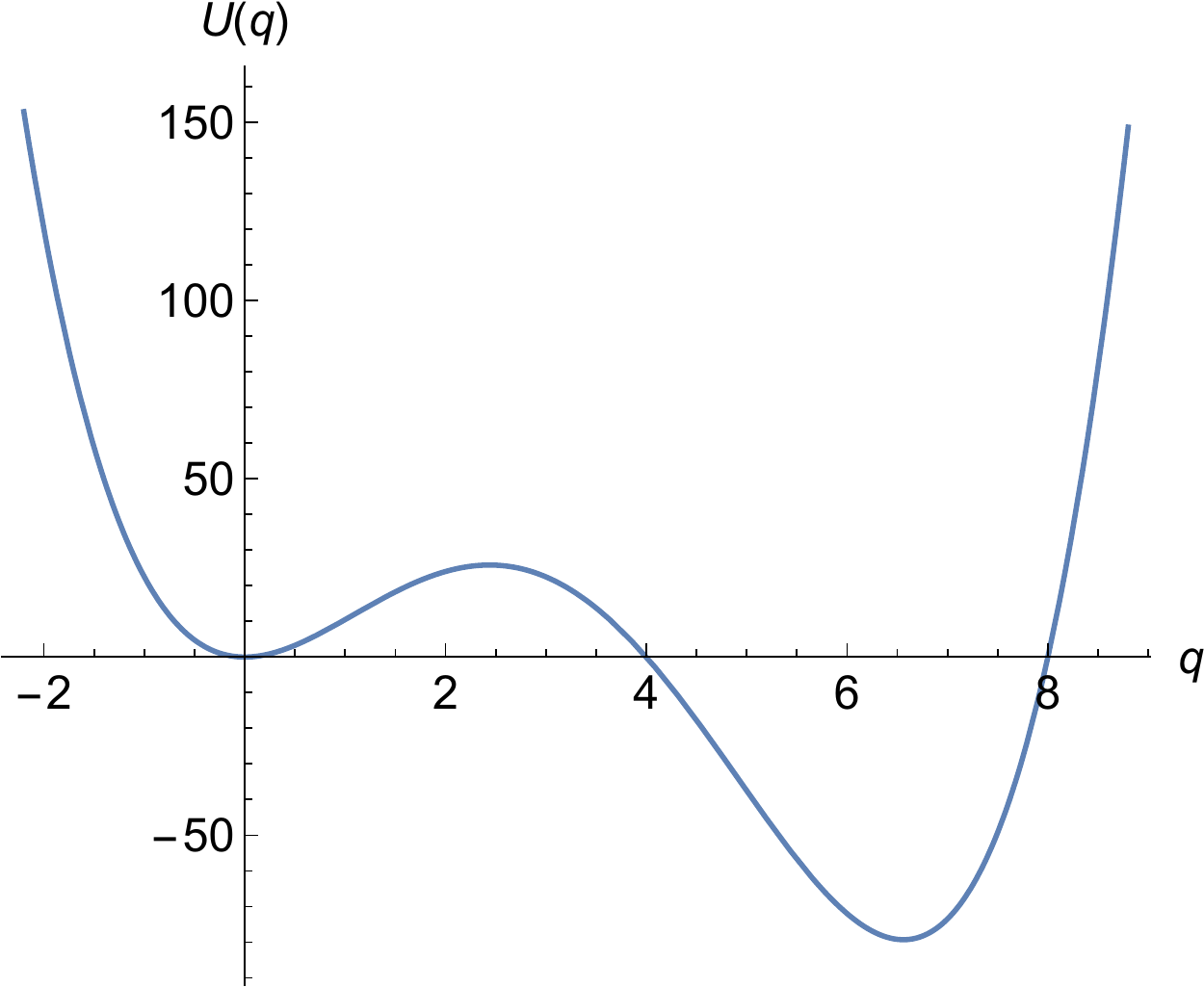}
		\caption{$U(q) = \frac{1}{2}q^2(q-4)(q-8)$}
		\label{fig:potential12}
	\end{subfigure}
	\caption{The potentials for two specific choices of parameters.}
	\label{fig:potentials}
\end{figure}
To compute thermal partition function $Z(\beta)$ we will use its path integral representation
\begin{eqnarray}
Z(\beta) = \int\mathcal{D}[q(t)] e^{- S(q)} ,
\end{eqnarray}
where the Euclidean action $S(q)$ is given by 
\begin{eqnarray}
S(q) = \int_{-\beta/2}^{\beta/2} dt \left[\frac{1}{2}(\dot{q}(t))^2 + U(q)\right] \label{origAction}
\end{eqnarray}
and the path integral is taken over periodic trajectories, such that
\begin{eqnarray}
q (-\beta/2) = q (\beta/2) .
\end{eqnarray}
We will be interested in the general potential of the form
\begin{eqnarray}
U (q) = \frac{1}{2} M q^2 (q-A) (q-B) = \frac{m}{2}q^2 + \frac{a}{3}q^3 + \frac{b}{4}q^4, 
\end{eqnarray}
where $M$, $A$ and $B$ ($m$, $a$ and $b$) are arbitrary parameters, such that $M, A, B > 0$ ($m > 0$, $a<0$, $b>0$). The two sets of parameters $M, A, B$ and $m, a ,b$
are related to each other via
\begin{eqnarray} 
M=\frac{b}{2},~~~A=\frac{-2a}{3b}\left(1-\sqrt{1-\frac{9bm}{2a^2}}\right),~~~B=\frac{-2a}{3b}\left(1+\sqrt{1-\frac{9bm}{2a^2}}\right),
\end{eqnarray}
This particular potential is the most general potential having qubic and quartic interaction terms in addition to the harmonic potential. The typical potentials generally considered in literature arise as its limiting cases. At the same time this particular potential is amenable to analytical treatment similar to the well known case of double well potential. In Fig. \ref{fig:potentials} we plotted potential for two specific choices of parameters.

\begin{figure}[h]
	\center{\includegraphics[width=0.8\textwidth]{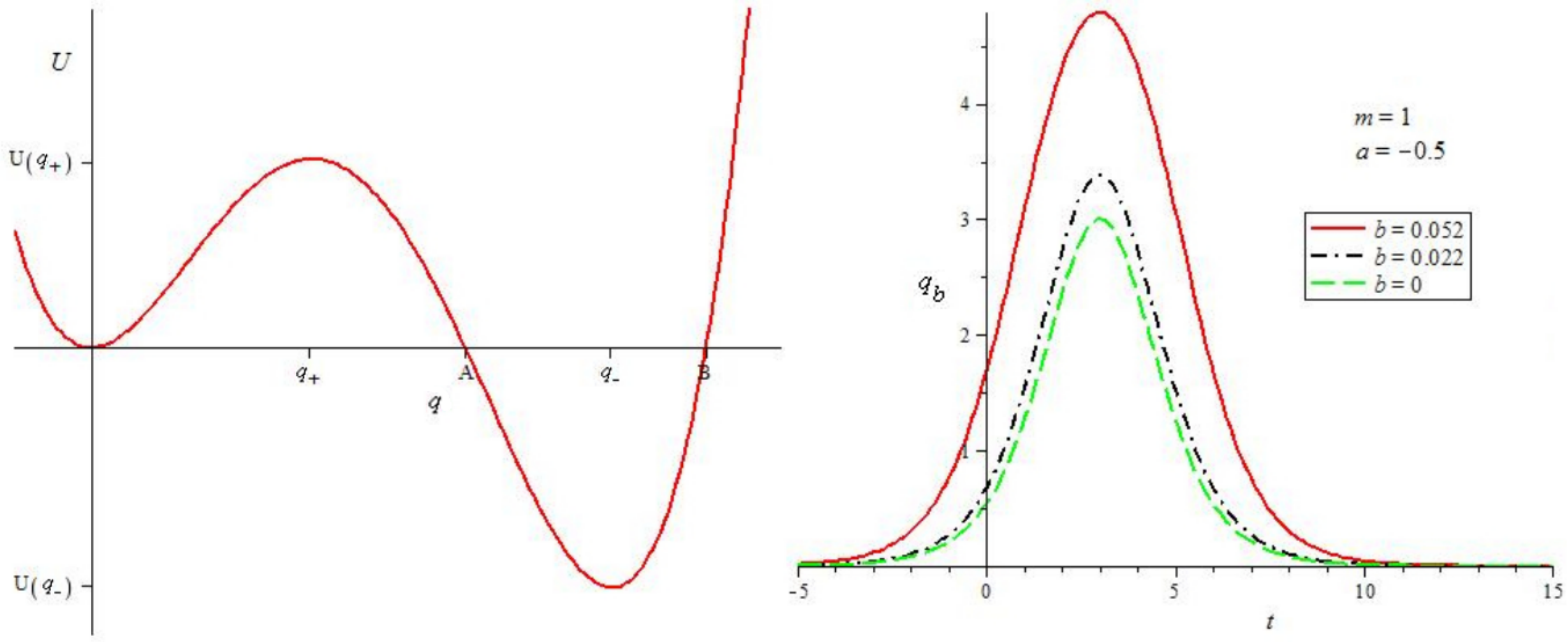}}
	\caption{$U(q)$ on the left and corresponding bounce solution on the right}
	\label{fig:phi3b-bounce}
\end{figure}

As we already mentioned in Introduction the real part of thermal partition function is given by the path integral around false vacuum, while the imaginary part is given by path integral around bounce solution\footnote{We refer interested reader for example to \cite{PrecisionDecayRates} for more details}. The bounce solution is the solution of classical equation of motion connecting the minima of the potential, which could be easily found using energy conservation for the inverted potential $V(q) = - U(q)$:
\begin{eqnarray}
\frac{1}{2} \dot{q}^2 - U(q) = 0 = E(\beta\to\infty)\, ,
\end{eqnarray}  
where we have stressed, that we are looking for $\beta = \infty$ solution corresponding to classical false vacuum energy. Integrating the above energy conservation condition we get the mentioned bounce solution parameterized by $t_0$ 
\begin{eqnarray}
q_{b} (t) = \frac{2 A B}{A + B + (B-A)\cosh (\sqrt{A B M}~(t-t_0))} .
\end{eqnarray}
Figure \ref{fig:phi3b-bounce} contains an example of bounce solution in the limit of small $b$, where $U(q_{\pm})$ and $q_{\pm}$ are given by
\begin{eqnarray} 
U(q_{\pm})=\frac{-1}{24b^3}\left(a^4-6a^2bm+6b^2m^2\mp a^4\left(1-\frac{4bm}{a^2}\right)^{\frac{3}{2}}\right)
\end{eqnarray}
\begin{eqnarray} 
q_{\pm}=-\frac{a}{2b}\left(1\mp\sqrt{1-\frac{4bm}{a^2}} \right)
\end{eqnarray}
Next, the bounce action is independent of $t_0$ and is given by
\begin{eqnarray}
S_{b} = \frac{1}{12}\sqrt{M}\left\{
\sqrt{A B}~(3 A^2 - 2 A B + 3 B^2) - 3 (A-B)^2 (A+B)\arctanh\left(
\frac{\sqrt{A}}{\sqrt{B}}
\right)
\right\}\, , \nonumber \\
\end{eqnarray}
where $\arctanh (x) = \frac{1}{2}\log\frac{1+x}{1-x}$. The original action (\ref{origAction}) may be further rewritten in terms of the deviation from the classical configuration, $\varphi (t)\equiv q(t) - q_{b}(t)$  as
\begin{eqnarray}
S &=& S_{b} + \frac{1}{2}\int_{-\beta/2}^{\beta/2} d t\, \varphi (t)\mathbb{D}\varphi (t) \nonumber \\
&& + \int_{-\beta/2}^{\beta/2} d t\, 
\left(
\frac{4 A B M}{A+B+(B-A)\cosh (T)} - \frac{1}{2} M (A+B)
\right)\varphi^3 (t) + \frac{1}{2} M\int_{-\beta/2}^{\beta/2} dt\, \varphi^4 (t)\, ,\nonumber \\ \label{lagrangian}
\end{eqnarray}
where
\begin{eqnarray}
\label{Dbounce}
\mathbb{D}\equiv -\frac{d^2}{d t^2} + A B M \left[
1 - \frac{6 B}{A + B + (B-A)\cosh (T)}
+ \frac{6 A ( (A-B)\cosh (T) - A + 3 B)}{(A+B+(B-A)\cosh (T))^2}
\right] \nonumber \\
\end{eqnarray}
and we have introduced the abbreviation $T = \sqrt{A B M}\, t$. While making transition to the deviations from bounce solution special care should be taken due to the existence of zero mode of operator $\mathbb{D}$. The latter is given by $\varphi_0 (t) = - S_{b}^{-1/2}\, \frac{d}{d t_0}\, q_{b}(t)$. To integrate over this zero mode we employ Faddeev-Popov-like trick as in \cite{AleinikovShuryak,Olejnik:1989id} and insert into our path integral the unity operator written as\footnote{See also \cite{InstantonsQCD,LecturesInstantons}.}
\begin{eqnarray}
1 = \int d t_0\, \delta (f(t_0))\, \frac{d f (t_0)}{d t_0}\, ,
\end{eqnarray}
where  
\begin{gather}
f (t_0) = \int \varphi (\tau) \varphi_0 (\tau)\, d\tau =\int (q (\tau) - q_{b} (\tau)) \varphi_0 (\tau)\, d\tau \, , \\
\frac{d f (t_0)}{d t_0} = S_{b}^{1/2} + \int\varphi (\tau)
\frac{d}{d t_0} \varphi_0 (\tau)\, d\tau \, .
\end{gather}
This way we get
\begin{eqnarray}
1 = \int d t_0 \left(
S_{b}^{1/2} + \int \varphi (\tau) \frac{d}{d t_0} \varphi_0 (\tau)\, d\tau
\right) \delta (c_0) , \label{tadpole-vertex}
\end{eqnarray}
where $c_0$ is the coefficient in front of our zero mode in the expansion of deviation $\varphi (\tau)$ in terms of eigenfunctions of $\mathbb{D}$ operator
\begin{equation}
\varphi (\tau) = \sum_n c_n \varphi_n (\tau)\, .
\end{equation}
We see, that this procedure gives us an extra tadpole vertex coming from integration measure in addition to those we get from lagrangian (\ref{lagrangian}). The presented procedure is known as a transition to collective coordinates. In the case under consideration the only collective coordinate present is given by $t_0$.

\subsection{Green function in the background of bounce solution}

The Green function at false vacuum is easy to find and it is given by the solution of corresponding Schr\"odinger equation: 
\begin{equation}
G_{FV}(t_1, t_2) = \frac{1}{2\sqrt{A B M}}\exp^{-\sqrt{A B M} ~|t_1-t_2|}\ 
\end{equation}
or 
\begin{equation}
G_{FV}(x, y) = \frac{1}{2\sqrt{A B M}}\frac{1 - |x-y|- x y }{1 + |x-y| - x y}\, ,
\end{equation}
where $x = \tanh \frac{\sqrt{A B M} t_1}{2}$ and $y = \tanh \frac{\sqrt{A B M} t_2}{2}$.

The determination of Green function in the background of bounce solution is more involved. As we have already seen in previous section the corresponding inverse operator $\mathbb{D}$ has zero mode. The inversion of the latter is consistently defined only on the subspace of functions orthogonal to this zero mode (Fredholm alternative). More concretely, the Green function we will be looking for is defined as\footnote{The normalization of wave functions to unity is assumed.} 
\begin{equation}
G_b (t_1, t_2) = \sum_{n, \lambda_n \neq 0} \frac{\varphi_n (t_1)\varphi_n (t_2)}{\lambda_n}\, ,
\end{equation}  
where $\mathbb{D} \varphi_n = \lambda_n\varphi_n$ and the summation goes over all modes except zero one. It is easy to check that the equation satisfied by this Green function is given by 
\begin{align}
&\mathbb{D}~G_b (t_1, t_2) = \delta (t_1 - t_2) - \varphi_0 (t_1) \varphi_0 (t_2)\, .
\end{align}
or
\begin{multline}
\frac{\partial}{\partial x}
\left(
(1-x^2)\frac{\partial G_b(x,y)}{\partial x}\right) -
\frac{4\left\{
	r^4 x^2 (3-2 x^2) + r^2 (3 x^4-8 x^2+3) + (3 x^2-2)	
	\right\}}{(1-x^2)(1-r^2 x^2)^2} G_b(x,y) \\
= - \frac{2}{\sqrt{A B M}}\delta (x-y) + \frac{\gamma x y (1-y^2)}{(1 - r^2 x^2)^2 (1 - r^2 y^2)^2}\, , 
\end{multline}
where $r = \sqrt{\frac{A}{B}}$ and
\begin{align}
\gamma = \frac{1}{\sqrt{A B M}}
\frac{48 r^5 (1-r^2)^2}{r (3-2 r^2+3 r^4)-3 (1-r^2)^2 (1+r^2)\tanh^{-1} (r)}
\end{align}
First solution of the homogeneous equation is given by the zero mode of operator $\mathbb{D}$ we found in previous section:
\begin{equation}
y_{hom, 1} (x) = \frac{x (1-x^2)}{(1-r^2 x^2)^2}\, . \label{hom1}
\end{equation}
The second solution may be found using the so-called reduction of order method. First,  from Abel's differential equation identity we get an equation for Wronskian of homogeneous solutions  $W (x)\equiv  y_{hom,1} (x)y_{hom,2}'(x)-y_{hom,1}'(x) y_{hom,2}(x)$:
\begin{equation}
\frac{d W (x)}{W (x)} = -d\log (1-x^2)\, .
\end{equation}
Integrating the latter together with subsequent first order differential equation (obtained from known expressions for Wronskian and first homogeneous solution) for $y_{hom, 2} (x)$ we get 
\begin{multline}
y_{hom,2} (x) = y_{hom, 1}(x)\int\frac{dx}{(1-x^2)  y_{hom, 1}^2(x)} \nonumber \\
= \frac{x (1-x^2)}{16 (1-r^2 x^2)^2} \Bigg\{
-\frac{16}{x} - 16 r^8 x + \frac{4 x (1-r^2)^4}{(1-x^2)^2}
+ \frac{2 (1-r^2)^3 (7+9 r^2) x}{1-x^2}  \nonumber \\
+ 3 (1-r^2)^2 (5+6 r^2+ 5 r^4)\log\frac{1+x}{1-x} \Bigg\}\, .
\end{multline}
The particular solution of nonhomogeneous solution is then found by the variation of constants and is given by
\begin{equation}
y_{nonhom} (x) = y_{hom, 2}(x) \int\frac{y_{hom ,1} (x) f(x)}{W (x)} dx - y_{hom ,1} (x) \int\frac{y_{hom ,2} (x) f(x)}{W (x)} dx\, ,
\end{equation}
where
\begin{equation}
f (x) = \frac{\gamma x y (1-y^2)}{(1-x^2)(1-r^2 x^2)^2 (1-r^2 y^2)^2}\, .
\end{equation}
\begin{figure}[th]
	\centering
	\begin{subfigure}{.5\textwidth}
		\centering
		\includegraphics[width=0.9\linewidth]{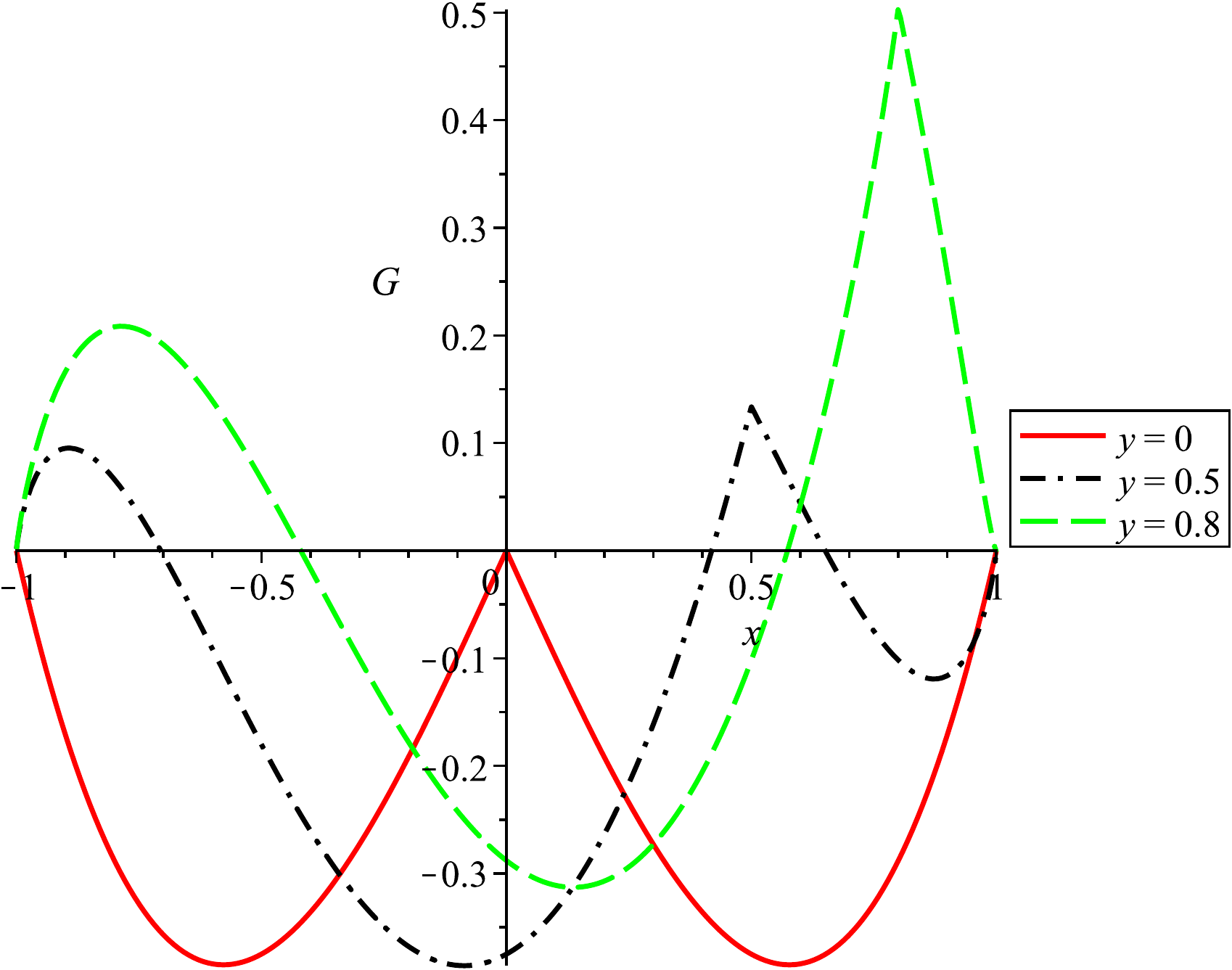}
		\caption{}
		\label{fig:greenfuncsphi3plot2D}
	\end{subfigure}%
	\begin{subfigure}{.5\textwidth}
		\centering
		\includegraphics[width=0.9\linewidth]{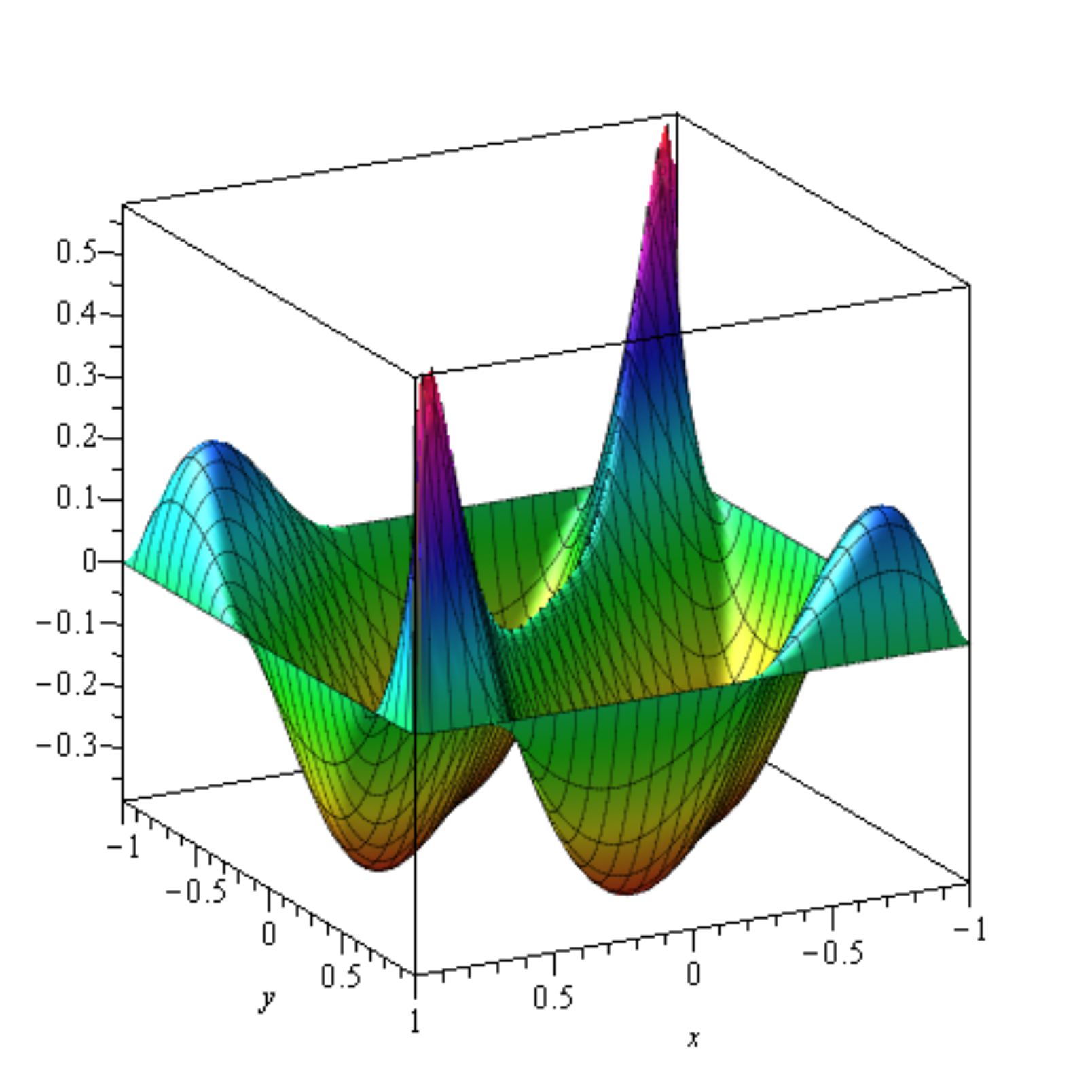}
		\caption{}
		\label{fig:greenfuncsphi3plot3D}
	\end{subfigure}
	\caption{Green function in the background of bounce solution for $m=1, b=0$: (a) - values of Green function at $y=0$ (solid line), $y=0.5$ (dot-dashed line), $y=0.8$ (dashed line), (b) - Green function as a function of $x, y$ variables}
	\label{fig:greenfuncsphi3}
\end{figure}
The integrals in the above expression could be evaluated in terms of polylogarithms. However, the expressions we get are quite lengthy and we put them in mathematica file accompanying this article. Here, we will present analytical expressions only for two particular cases. In the limit\footnote{One should take a limit and not just set $b$ to zero.} $b\to 0$ we get 
\begin{eqnarray} 
y_{nonhom} (x, y) = \frac{3xy(1-y^2)}{4}\left(1+\frac{1}{3(1-x^2)}\right) \, 
\end{eqnarray}
and in the case $r = \sqrt{\frac{A}{B}} = \frac{1}{\sqrt{2}}$ the particular solution takes the form:
\begin{multline}
y_{nonhom} (x, y) = y_0 (x, y)\Bigg\{
\frac{183-89x^2-72x^4}{96 (x^2-1)^2} - \frac{37}{256} (44+9\sqrt{2}\log (3-2\sqrt{2}))\log (1-x) \\
+\frac{37}{256}(-44+9\sqrt{2}\log (3+2\sqrt{2}))\log (1+x) + \frac{407}{64}\log (2-x^2) \\
- \frac{3 (128 - 273 x^2 + 135 x^4 + 8 x^6)}{64\sqrt{2} x (x^2-1)^2}\log\left(\frac{2-\sqrt{2}x}{2+\sqrt{2}x}\right)
+ 2\log\left(\frac{x^2-1}{x^2-2}\right) \\
- \frac{333}{128\sqrt{2}}\Big[
\Li2 ((1-\sqrt{2})(x-1)) - \Li2 ((1+\sqrt{2})(x-1))
+ \Li2 ((\sqrt{2}-1)(1+x)) - \Li2 (-(1+\sqrt{2})(1+x))\Big]\Bigg\}\, ,
\end{multline}
where
\begin{equation}
y_0 (x, y) = \frac{\gamma \big|_{r=1/\sqrt{2}}~ x y (1-x^2)(1-y^2)}{(2-x^2)^2 (2-y^2)^2}\, .
\end{equation}
The Green function in the background of bounce solution is then given by
\begin{equation}
G(x,y) = C_1 (y) y_{hom,1}(x) + C_2 (y) y_{hom,2}(x) + y_{nonhom}(x,y)\, 
\end{equation}
with $C_1(y)$ and $C_2(y)$ are some arbitrary functions of variable $y$. The latter are determined from the conditions:
\begin{enumerate}
	\item $G(x,y)$ is finite for $x\to\pm 1$ ($t_1\to\pm\infty$);
	\item $G(x,y)$ is continuous at $x=y$ ($t_1 = t_2$);
	\item $G(x,y)$ is orthogonal to the zero mode $y_{hom,1}(x)$ (\ref{hom1}) (we accounted for the Jacobian of transition from $t_1$ to $x$, $J = \frac{2}{1-x^2}$): 
	\begin{equation}
	\int_{-1}^{1} \frac{d x}{1-x^2} y_{hom, 1}(x) G (x, y) = 0\, . \label{zero-mode-orthogonality}
	\end{equation}
\end{enumerate} 
\begin{figure}[h]
	\centering
	\begin{subfigure}{.5\textwidth}
		\centering
		\includegraphics[width=0.9\linewidth]{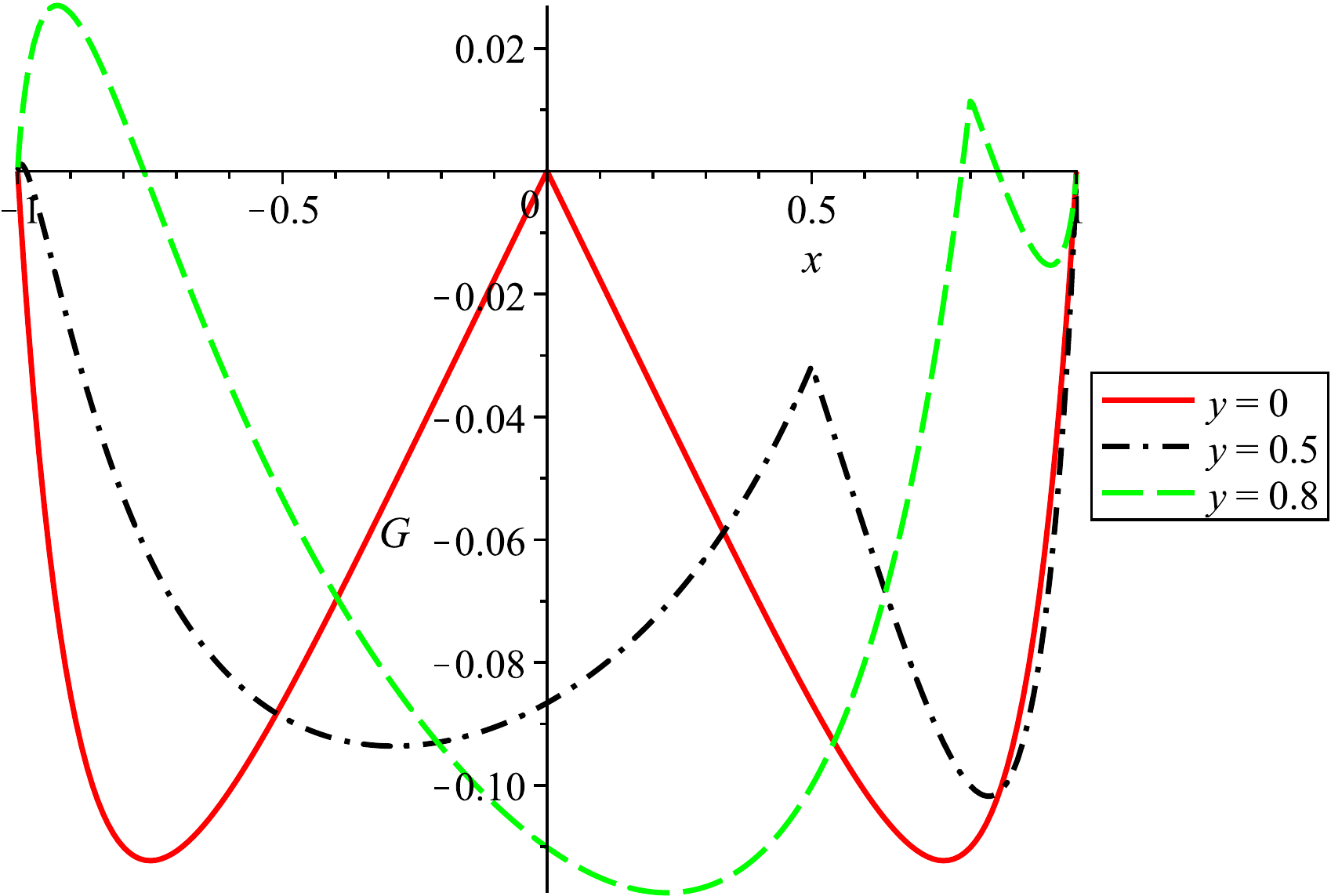}
		\caption{}
		\label{fig:greenfuncsAB12plot2D}
	\end{subfigure}%
	\begin{subfigure}{.5\textwidth}
		\centering
		\includegraphics[width=0.9\linewidth]{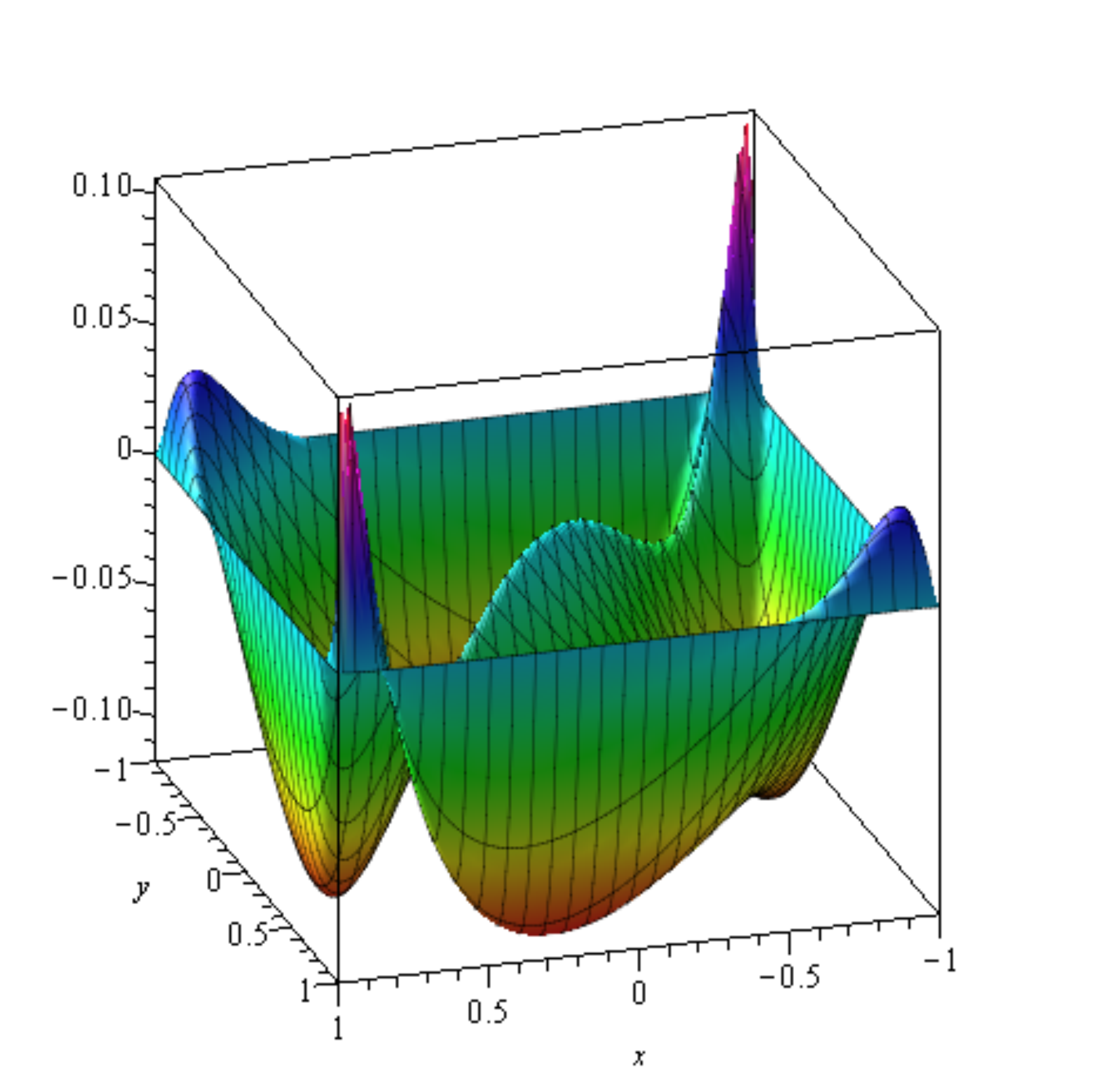}
		\caption{}
		\label{fig:greenfuncsAB12plot3D}
	\end{subfigure}
	\caption{Green function in the background of bounce solution for $M=1, A=4, B=8$: (a) - values of Green function at $y=0$ (solid line), $y=0.5$ (dot-dashed line), $y=0.8$ (dashed line), (b) - Green function as a function of $x, y$ variables}
	\label{fig:greenfuncsAB12}
\end{figure}
Using the first two conditions together with the fact that Green function is symmetric with respect to the exchange of variables $x\leftrightarrow y$ we were able to determine Green function analytically up to one unknown constant. The latter could be found using third condition numerically. The resulting expression for Green function is too lengthy and could be found in accompanying mathematica file. In the limit $b\to 0$ the analytical expression is quite short and the corresponding Green function is given by ($\frac{bm}{a^2}\ll 1$ or $\frac{A}{B}\ll 1$)
\begin{eqnarray} 
G(x,y) = G_0(x,y) + b G_1(x,y) + \mathcal{O}(b^2)\, ,
\end{eqnarray}
where
\begin{multline}
\label{GreenQ0} 
G_0(x,y)=\frac{1}{2 m^{1/2}}\Bigg[g_0(x,y)\Big[(x^2+y^2)(6xy-4)+3xy(4-5xy) \\
-\frac{1}{4}|x-y|(8-55xy+15x^2y^2+8(x^2+y^2))\Big] \\ 
+\frac{15}{8}xy(1-x^2)(1-y^2)\left(\log{(g_0(x,y))}-\frac{31}{15}\right)
\Bigg]\, , 
\end{multline}
\begin{multline}
G_1(x,y)=\frac{9m^{1/2}}{16a^2}\Bigg[3xy(x^2+y^2-x^4-y^4)-xy\left(\frac{(1-y^2)^2}{1-x^2} +\frac{(1-x^2)^2}{1-y^2}\right) \\ +\frac{|x-y|}{2}\bigg\{ xy(6-5(x^2+y^2))+
+26x^2y^2-10(x^2+y^2-x^4-y^4)-15x^2y^2(x^2+y^2)\\+2(x+y)\left(\frac{x(1-x^2)}{1-y^2}+\frac{y(1-y^2)}{1-x^2}\right)\bigg\}\nonumber \\
+\frac{3}{28}xy(1-x^2)(1-y^2)\bigg\{35(x^2+y^2)\left(\log{(g_0(x,y))}-\frac{217}{105}\right)-30\log{(1-|x-y|-xy)}\nonumber \\
-2\log{(1+|x-y|-xy)}-\frac{242}{5}+64\log{2}\bigg\}\Bigg]
\end{multline}
and 
\begin{equation}
 \label{go} 
 g_0(x,y)=\frac{1-|x-y|-xy}{1+|x-y|-xy}
\end{equation}
Finally, figures \ref{fig:greenfuncsphi3} and \ref{fig:greenfuncsAB12} contain plots of Green functions in the background of bounce solution for two sets of parameters $m=1, b=0$ and $M=1, A=4, B=8$.

\subsection{One loop expression}

At one loop the real and imaginary parts of our thermal partition function are given by\footnote{See for example \cite{InstantonsLargeN,PrecisionDecayRates}.} ($Z_{FV}$ is the partition function evaluated by the expansion at false vacuum ):
\begin{align}
\text{Re} Z &= Z_{FV} (\beta) \\
\text{Im} Z &= \frac{1}{2 i} Z_{FV} (\beta) \lim_{\beta\to\infty} \left[-\frac{\det '\mathbb{D}}{\det\mathbb{D}_{FV}}\right]^{-1/2} \frac{\beta S_b^{1/2}}{\sqrt{2\pi}} e^{-S_b}
\end{align}
so, that for imaginary part of energy we get 
\begin{equation}
\text{Im} E = \frac{S_b^{1/2}}{2\sqrt{2\pi}}\lim_{\beta\to\infty}  \left[-\frac{\det '\mathbb{D}}{\det\mathbb{D}_{FV}}\right]^{-1/2}  e^{-S_b}\, .
\end{equation}
The easy way to derive an expression for the ratio of functional determinants is to use the formalism of  Gel'fand and Yaglom \cite{GenfaldYaglom}, see also \cite{InstantonsLargeN,DunneFunctionalDeterminants}. Within the latter, given a Sch\"odinger operator defined on the interval $t\in [-\frac{\beta}{2},\frac{\beta}{2}]$ with eigenfunctions satisfying Dirichlet\footnote{In the zero temperature case considered here it is enough to consider Dirichlet boundary conditions.} boundary conditions 
\begin{equation}
L\, \psi (t) = 
\left[
-\partial_t^2 + U(t)
\right]\psi (t) = k^2\psi (t), \quad \psi (-\frac{\beta}{2}) = \psi (\frac{\beta}{2}) = 0
\end{equation} 
its determinant is found as a solution of the auxiliary problem
\begin{equation}
\left[
-\partial_t^2 + U(t)
\right]\phi (t)  = 0,\quad \phi (-\frac{\beta}{2}) = 0\, ,\, \dot{\phi} (-\frac{\beta}{2}) = 1
\end{equation}
so, that
\begin{equation}
\text{det}\left[
-\partial_t^2 + U (t)
\right]  = \phi (\frac{\beta}{2})
\end{equation}
In general the above determinants diverge, so one generally considers their ratio
\begin{equation}
\frac{\text{det} L_1}{\text{det} L_2} = \frac{\phi_1 (\frac{\beta}{2})}{\phi_2 (\frac{\beta}{2})}
\end{equation} 
This result could be straightforwardly obtained using contour integration technique of Kirsten and McKane \cite{KirstenMcKane1,KirstenMcKane2}. Indeed, writing the eigenvalue problem as
\begin{equation}
(L_j - k^2) u_{j,k} (t) = 0,\quad u_{j,k} (-\frac{\beta}{2}) = 0\, ,\, \dot{u}_{j,k} (-\frac{\beta}{2}) = 1
\end{equation}   
and noting that if $u_{j,k} (t)$ satisfies the Dirichlet boundary conditions on both sides of the interval, i.e. if also
\begin{equation}
u_{j,k} (\frac{\beta}{2}) = 0
\end{equation}
then  $k^2$ becomes the eigenvalue of $L_j$ and the sum of chosen powers of all eigenvalues of the latter could be conveniently represented by the following contour integral
\begin{equation}
\zeta_{L_j} (s) = \sum_{n=1}^{\infty}\frac{1}{k_n^{2 s}} = \frac{1}{2\pi i}\int_{\gamma} k^{-2 s}\frac{d}{d k} \ln u_{j,k} (\frac{\beta}{2})\, d k \, , 
\end{equation}
where the contour runs counterclockwise and we used the fact, that in the vicinity of eigenvalue $k_n$
\begin{equation}
\frac{d}{d k}\ln u_{j,k} (\frac{\beta}{2}) \approx \frac{1}{k - k_n}\, .
\end{equation}  
The logarithmic derivative converges\footnote{At large $k$ the potential in Scr\"odinger operator could be neglected and we have $u_{j,k}\approx \sin (k x)\left(1+\mathcal{O}(k^{-1})\right)$.} as $|k|\to\infty$ and the integration contour could be deformed as we wish. Deforming the latter to imaginary axis we get
\begin{equation}
\zeta_{L_1} (s) - \zeta_{L_2} (s) =\frac{\sin (\pi s)}{\pi}\int_0^{\infty} k^{-2 s} \frac{d}{d k}\ln 
\frac{u_{1,i k}(\frac{\beta}{2})}{u_{2,i k}(\frac{\beta}{2})}\, d k
\end{equation}
Recalling now that the operator determinants could be expressed as exponentials  of the introduced zeta functions derivatives at zero,  the determinant ratio is given by
\begin{equation}
\frac{\det L_1}{\det L_2} = e^{\zeta_{L_2}^{'} (0) - \zeta_{L_1}^{'} (0)} = \frac{u_{1,0} (\frac{\beta}{2})}{u_{2,0} (\frac{\beta}{2})}\, .
\end{equation}
The presented derivation is valid if the considered operators do not contain zero modes. In our case $L_1$ does actually have a zero mode due to time translation invariance and the above contour deformation is ill defined as $u_{1,k}(\frac{\beta}{2})$ is zero at $k = 0$. To overcome this difficulty we need to know the behavior of $u_{1,k} (\frac{\beta}{2})$ for small $k$ to eliminate the pole in integrand. Integrating by parts the left hand side of 
\begin{equation}
\int_{-\frac{\beta}{2}}^{\frac{\beta}{2}} dt\, u_{1,0} (t)^{*} L_1 u_{1,k} (t) =  k^2\int_{-\frac{\beta}{2}}^{\frac{\beta}{2}} dt\, u_{1,0} (t)^{*} u_{1,k} (t)\equiv k^2 \la u_{1,0} | u_{1,k}\ra 
\end{equation}
gives 
\begin{equation}
\left[\dot{u}_{1,0}(t)^{*} u_{1,k} (t) - \dot{u}_{1,k}(t)^{*} u_{1,0} (t)\right]_{-\frac{\beta}{2}}^{\frac{\beta}{2}} + \int_{-\frac{\beta}{2}}^{\frac{\beta}{2}} dt\, u_{1,k} (t) (L_1 u_{1,0} (t))^{*} = k^2 \la u_{1,0} | u_{1,k}\ra\, .
\end{equation}
Using the boundary conditions for $u_{1,0}(t)$ and $u_{1,k}(t)$ we get\footnote{We dropped the $*$ as we are dealing with real solutions only.}
\begin{equation}
u_{1,k}(\frac{\beta}{2}) = \frac{k^2\la u_{1,0}| u_{1,k}\ra}{\dot{u}_{1,0}(\frac{\beta}{2})}\equiv -k^2 f_{1,k}.
\end{equation}  
It is easy to see that due to the orthogonality of eigenfunctions $f_{1,k}$  vanishes for all values of $k^2$ except at $k = 0$, where it remains non-zero. A function which  behaves as $u_{1,k}(\frac{\beta}{2})$ for large $k$ but is non-zero for $k=0$ is given by $(1-k^2)f_{1,k}$, so the contour integral associated with the determinant of $L_1$ is given by  
\begin{equation}
\frac{1}{2\pi i}\int_{\gamma} k^{-2 s}\frac{d}{d k}\ln (1-k^2) f_{1,k} d k = \zeta_{L_1} (s) + \frac{1}{2\pi i}\int_{\gamma} k^{-2 s} \frac{d}{d k}\ln (1-k^2) d k\, ,
\end{equation}
where the zero mode from $\zeta_{L_1} (s)$  was omitted. The second integral on the right hand side is easily taken by a residue at simple pole $k=1$ and we get  
\begin{equation}
\zeta_{L_1} (s) - \zeta_{L_2} (s) = \frac{\sin (\pi s)}{\pi} \int_0^{\infty} k^{-2 s} \frac{d}{d k}\ln 
\left[
\frac{(1+k^2) f_{1, i k}}{u_{2, i k} (\frac{\beta}{2})}
\right] - 1 \, .
\end{equation}
The expression for the corresponding determinant ratio is then given by
\begin{equation}
\frac{\det ' L_1}{\det L_2} = e^{\zeta_{L_2}^{'} (0) - \zeta_{L_1}^{'} (0)} = \frac{f_{1,0}}{u_{2,0} (\frac{\beta}{2})} = -\frac{\la u_{1,0}| u_{1,0}\ra}{\dot{u}_{1,0} (\frac{\beta}{2}) u_{2,0} (\frac{\beta}{2})} \, ,
\end{equation}
where $\det '$ denotes the determinant with zero eigenvalue omitted. The expression for $u_{1,0} (x)$ function
in the case of functional determinant around bounce solution could be easily found using the methods presented in previous section and is given by
\begin{equation}
u_{1,0}(t) = \dot{q}_{b} (-\beta/2) \dot{q}_{b} (t) \int_{-\frac{\beta}{2}}^t \frac{d t'}{(\dot{q}_b (t'))^2} . \label{u10t}
\end{equation}
Here, we assumed that the parameter $t_0$ of the bounce solution was chosen, so that $\ddot{q}_b (-\frac{\beta}{2}) = 0$. This result could be easily generalized for more complicated boundary conditions both with and without zero modes \cite{KirstenMcKane1,KirstenMcKane2}. Finally, in our particular case we get
\begin{equation}
\lim_{\beta\to\infty}\frac{\det ' L_b}{\det L_{FV}} = 
\lim_{\beta\to\infty}\frac{\det '\mathbb{D}}{\det\mathbb{D}_{FV}} = -\frac{(1-r^2)^2}{32 r^7 B^5 M^{3/2}} S_b\, ,
\end{equation}
so that the imaginary part of energy is given by 
\begin{equation}
\text{Im} E = \frac{2 A B}{B-A}\frac{M^{3/4} A^{3/4} B^{3/4}}{\sqrt{\pi}} e^{-S_b}\, .
\end{equation}
We have checked that the same expression is reproduced using formula presented in \cite{KleinertChervyakov1,KleinertChervyakov2}. Moreover, it could be also obtained using slight modification of the derivation presented in \cite{InstantonsLargeN} for the case of $U''(0) = \omega^2\neq 1$.  The above determinant ratio in this case is given by
\begin{eqnarray}
\lim_{\beta\to\infty}\frac{\det '\mathbb{D}}{\det\mathbb{D}_{FV}} = -\frac{S_b}{2\omega q_0^2}\exp{ \left( -2\int\limits_0^{q_0}dx\left[\frac{\omega}{\sqrt{2V(x)}}-\frac{1}{ x} \right] \right)}\, ,
\end{eqnarray}
where $q_0$ is the zero of potential and turning point for the bounce solution. In our case $q_0 = A$. The corresponding imaginary part of energy is then found with 
\begin{eqnarray}
\text{Im} E = \frac{\omega^{\frac{3}{2}}q_0}{2\sqrt{\pi}}\exp{ \left(\int\limits_0^{q_0}dx\left[\frac{\omega}{\sqrt{2V(x)}}-\frac{1}{ x} \right] \right)}e^{-S_b}
\end{eqnarray}
\begin{figure}[h]
	\center{\includegraphics[width=0.8\textwidth]{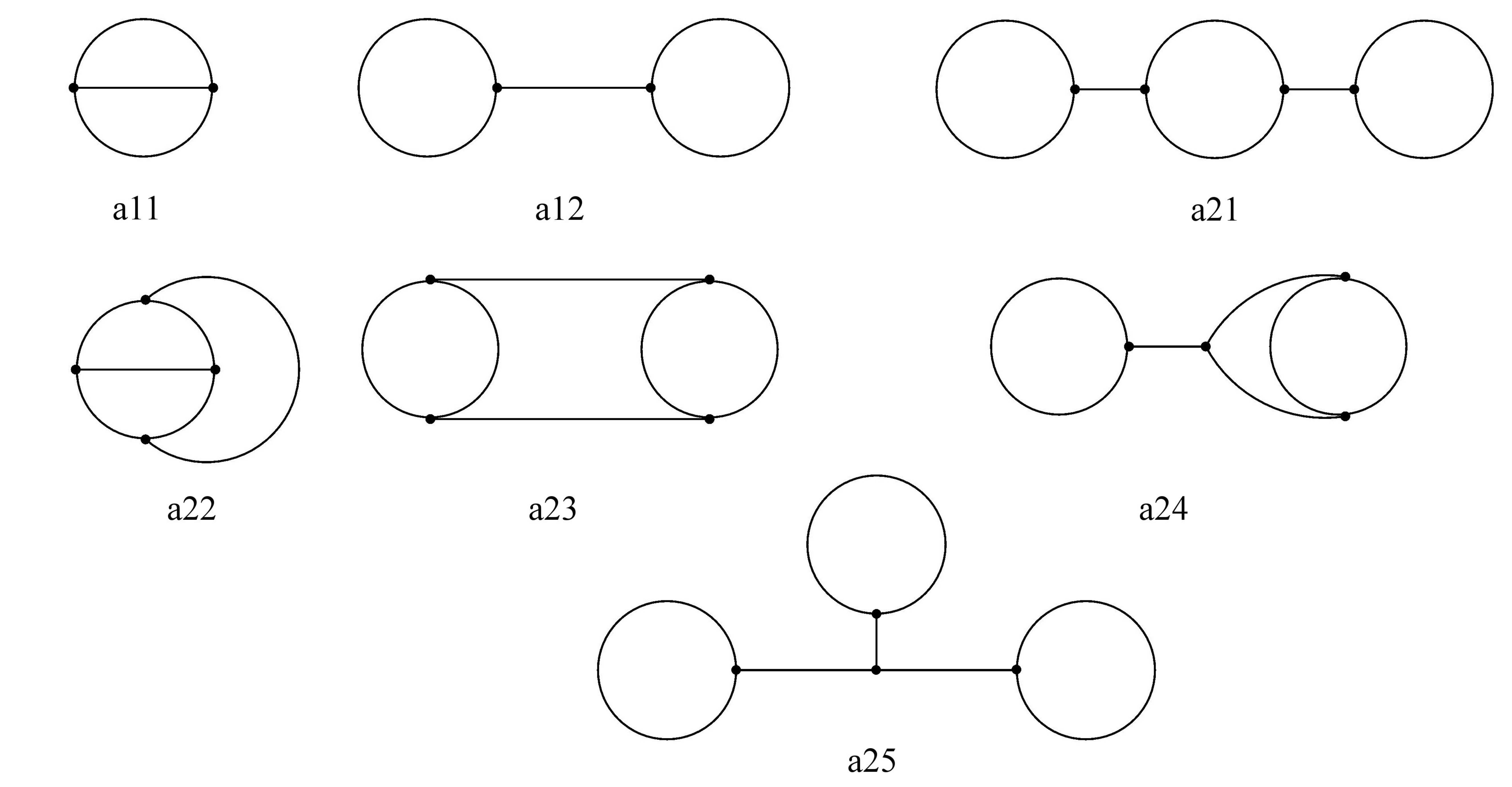}}
	\caption{Two and tree loop Feynman diagrams with only cubic interaction vertexes.}
    \label{FDcubic}
\end{figure}
\begin{figure}[h]
	\center{\includegraphics[width=0.8\textwidth]{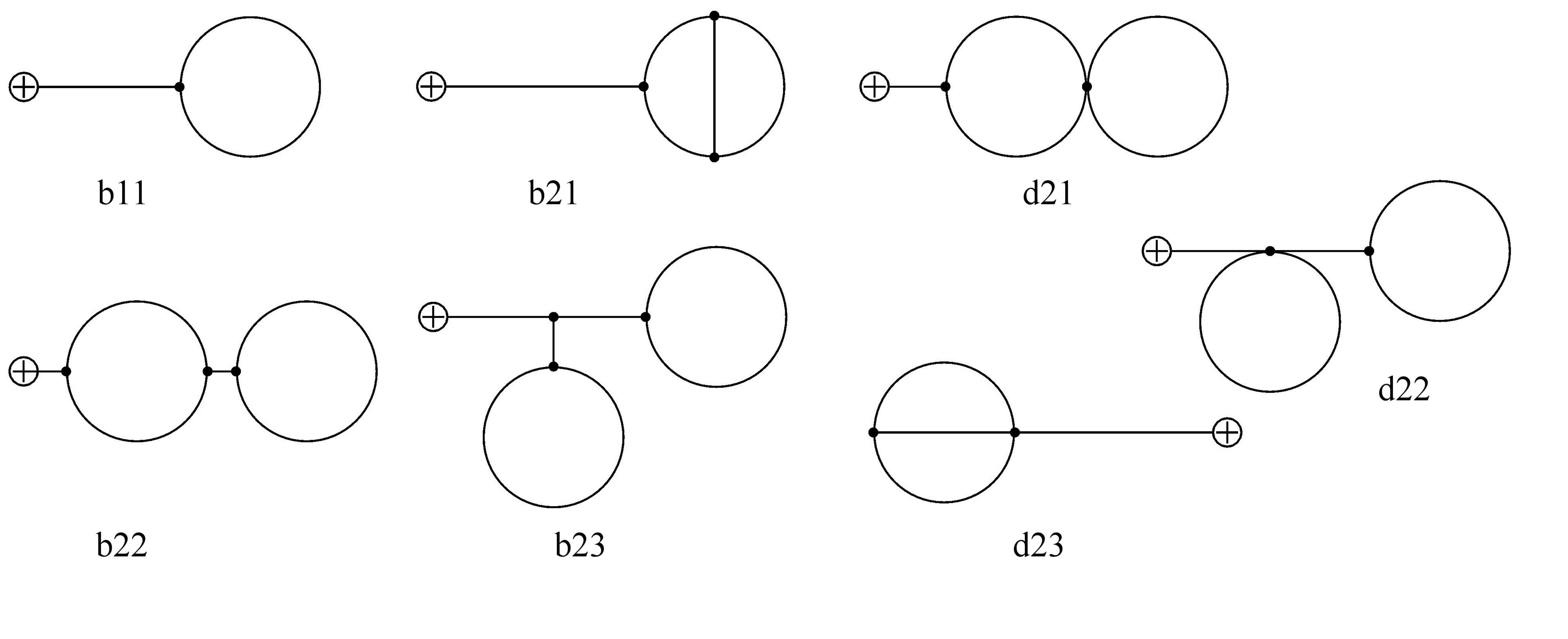}}
	\caption{Two and tree loop Feynman diagrams with tadpole vertexes.}
	\label{FDtadpole}
\end{figure}

\subsection{Non-gaussian effects}

To calculate higher order corrections we are following \cite{AleinikovShuryak,Olejnik:1989id,Shuryak1,Shuryak2,Shuryak3} and use Feynman diagram technique on top of false vacuum and bounce solution. The first is used to calculate higher order contributions to real part of thermal partition function, while the latter for higher order radiative corrections to its imaginary part. The corresponding Feynman rules are easy to derive and they are given by
\begin{align}
& V_{3, FV} = 3 M (A+B)\, ,\quad & V_{4, FV} = -12 M\, , \nonumber \\ 
& V_{3, b} = 3 M\left(
A+B - \frac{4 A B (1-x^2)}{B-A x^2}
\right)\, , \quad & V_{4, b} = -12 M\, , \nonumber \\
& V_{tad} = -\frac{A^2 (A-B) B^2 M (1-x^2) \left(B - 3(B-A)x^2 - A x^4\right)}{2 S_b (B-A x^2)^3} \, .
\end{align}
\begin{figure}[h]
	\center{\includegraphics[width=0.8\textwidth]{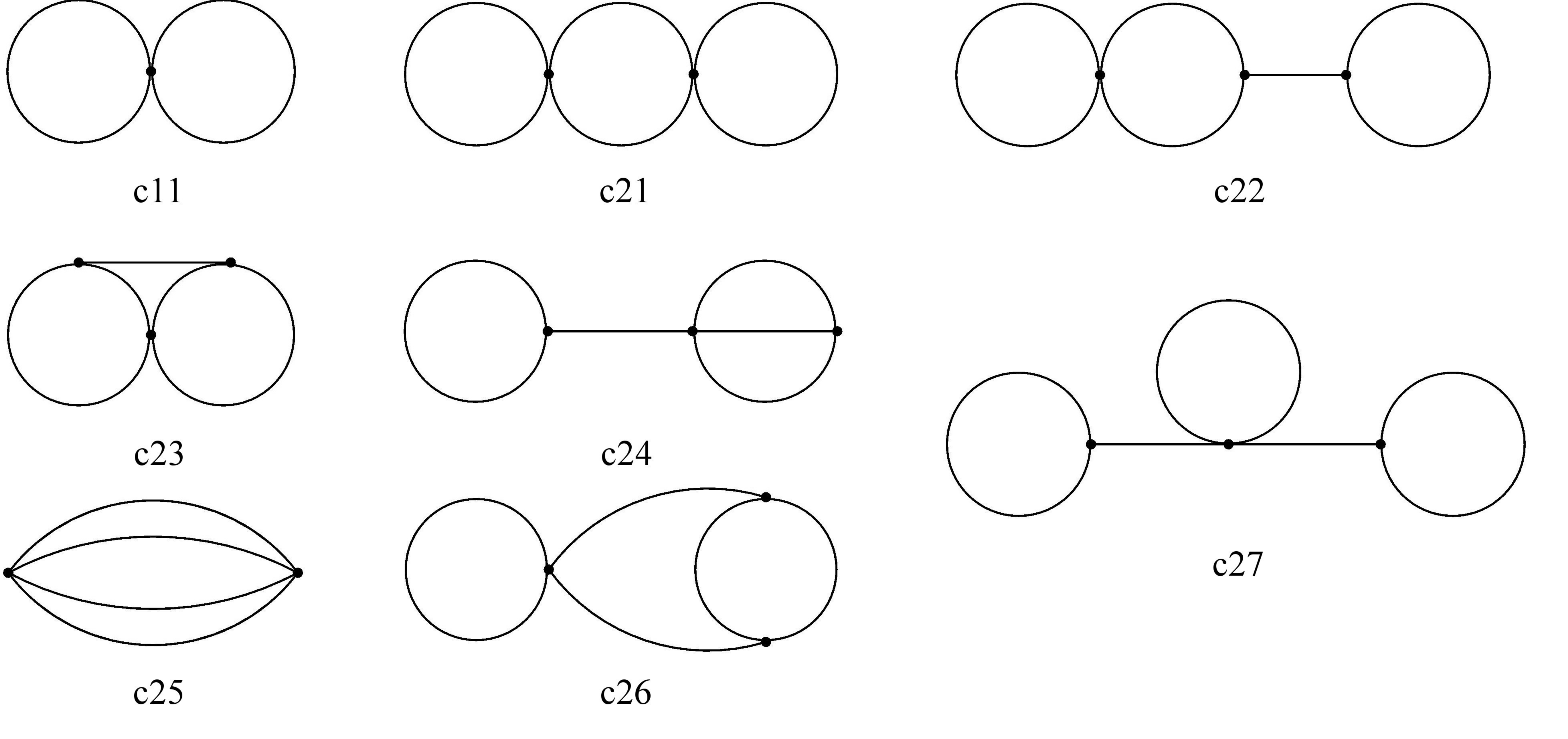}}
	\caption{Two and tree loop Feynman diagrams with quartic in addition to cubic interaction.}
	\label{FDquartic}
\end{figure}

Note, that Feynman diagrams on top of bounce solution may contain single tadpole vertex coming from the Jacobian of transition to collective coordinate (\ref{tadpole-vertex}). Such diagrams appear only for the expansion on top of bounce solution and are absent in the case of false vacuum.  The Green functions for the cases of false vacuum and bounce solution were given in previous subsection.  Figures \ref{FDcubic}, \ref{FDtadpole} and \ref{FDquartic} contain two and tree loop diagrams  containing only cubic vertexes,  tadpole vertex and extra quartic vertexes correspondingly. The diagram expressions with account for symmetry factors are constructed as it is usually done in quantum field theory. For example, the expression for Feynman diagram $b22$ (see Fig. \ref{FDtadpole}) is given by:     
\begin{eqnarray} 
I_{b22}=\frac{1}{4}\int\limits_{-1}^1dx\int\limits_{-1}^1dy\int\limits_{-1}^1dz\int\limits_{-1}^1dwJ_4
V_{tad}(x)V_{3,b}(y)V_{3,b}(z)V_{3,b}(w)G(x,y)G^2(y,z)G(z,w)G(w,w) \nonumber \\
\end{eqnarray}
where $J_n=\prod\limits_{i=1}^n2/(1-x_i^2)$ is the Jacobian of transition from $t_1,t_2,...t_n$ to $x,y,...,w$ variables.
\begin{table}[t]
	\caption{\label{phi3corrections} Contributions of all two ant three loop diagrams in the case of  $U(q)=\frac{1}{2}q^2+\frac{a}{3}q^3,$ potential. Here, $s$ is the symmetry factor, $i$ - diagram label and $I_i$ - the diagram numerical value with absolute error ( $b11$ is calculated exactly).}
	\centering
	\begin{tabular}{|c|c|c|}
		\hline \multirow{1}{2cm}{~~~~~~~$i$}&\multirow{1}{3cm}{~~~~~~~$s$}&\multirow{1}{5cm}{$~~~~~~I_i^0$} \\
		\hline
		\multirow{1}{2cm}{~~~~a11}&\multirow{1}{3cm}{~~~~~$1/12$}&\multirow{1}{5cm}{$-0.162143\pm 1.97*10^{-6}$} \\
		\hline
		\multirow{1}{2cm}{~~~~a12}&\multirow{1}{3cm}{~~~~~$1/8$}&\multirow{1}{5cm}{$0.0847623\pm 7.32*10^{-6}$} \\
		\hline
		\multirow{1}{2cm}{~~~~b11}&\multirow{1}{3cm}{~~~~~$1/2$}&\multirow{1}{5cm}{$-559/420$} \\
		\hline
		\multirow{1}{2cm}{~~~~a21}&\multirow{1}{3cm}{~~~~~$1/16$}&\multirow{1}{5cm}{$0.4497\pm1.5*10^{-4}$} \\
		\hline
		\multirow{1}{2cm}{~~~~a22}&\multirow{1}{3cm}{~~~~~$1/24$}&\multirow{1}{5cm}{$-0.01245\pm1.5*10^{-5}$} \\
		\hline
		\multirow{1}{2cm}{~~~~a23}&\multirow{1}{3cm}{~~~~~$1/16$}&\multirow{1}{5cm}{$0.100874\pm4*10^{-5}$} \\
		\hline 
		\multirow{1}{2cm}{~~~~a24}&\multirow{1}{3cm}{~~~~~$1/8$}&\multirow{1}{5cm}{$0.03801\pm1.6*10^{-4}$} \\
		\hline 
		\multirow{1}{2cm}{~~~~a25}&\multirow{1}{3cm}{~~~~~$1/48$}&\multirow{1}{5cm}{$0.3942\pm0.0042$} \\
		\hline   
		\multirow{1}{2cm}{~~~~b21}&\multirow{1}{3cm}{~~~~~$1/4$}&\multirow{1}{5cm}{$-0.4297\pm6.2*10^{-4}$} \\
		\hline 
		\multirow{1}{2cm}{~~~~b22}&\multirow{1}{3cm}{~~~~~$1/4$}&\multirow{1}{5cm}{$-0.9156\pm1.2*10^{-4}$} \\
		\hline      
		\multirow{1}{2cm}{~~~~b23}&\multirow{1}{3cm}{~~~~~$1/8$}&\multirow{1}{5cm}{$-1.2745\pm0.0016$} \\
		\hline                    
	\end{tabular}
\end{table}

Diagrams without tadpole vertexes are present both in corrections to real and imaginary parts of thermal partition function. As the imaginary part of energy we are interested in is defined by their ratio it is convenient to combine these contributions, so that for example the difference of corresponding diagrams $c24$ for bounce and false vacuum solutions is given by:     
\begin{eqnarray} 
I_{c24}=\frac{1}{12}\int\limits_{-1}^1dx\int\limits_{-1}^1dy\int\limits_{-1}^1dzJ_3
\big(V_{3,b}(x)V_{4,b}(y)V_{3,b}(z)G(x,x)G(x,y)G^3(y,z) - \nonumber \\ 
-V_{3,FV}(x)V_{4,FV}(y)V_{3,FV}(z)G_{FV}(x,x)G_{FV}(x,y)G_{FV}^3(y,z)\big)
\end{eqnarray}
To calculate two and three loop corrections to the false vacuum decay rate for arbitrary values of parameters $A, B, M$ one may use the mathematica notebook accompanying this article, where required numerical integrations are performed with the help of CUBA library \cite{CUBAlibrary}. Here we will present results only for two particular cases used throughout this paper. In the limit $b\to 0$ the imaginary part of energy up to three-loop accuracy is given by 
\begin{eqnarray} 
\text{Im}\, E =\frac{2AB}{B-A}\frac{M^{3/4} A^{3/4} B^{3/4}}{\sqrt{\pi}}e^{-S_b}\left\{1+\frac{I_2^0+\frac{b}{a^2}I_2^1+O\left(\frac{b^2}{a^4}\right)}{S_b}+\frac{I_3^0+\frac{b}{a^2}I_3^1+O\left(\frac{b^2}{a^4}\right)}{S_b^2}+\mathcal{O}(1/S_b^3)\right\} \nonumber \\
\end{eqnarray}
where in this case $S_b=\frac{6}{5a^2} + \frac{4131}{560}\frac{b}{a^4} + \mathcal{O}(b^2)$ and $I_2^i$ are two loop and $I_3^i$ - three loop contributions (see Tables  \ref{phi3corrections} and \ref{linearcontr}): 
\begin{eqnarray} 
I_2^0=I_{a11}^0+I_{a12}^0+I_{b11}^0,~~~~I_{3}^0=\sum\limits_{k=1}^{5}I_{a2k}^0+\sum\limits_{k=1}^
{3}I_{b2k}^0+I_{2contr}^0\, ,
\end{eqnarray}
with
\begin{eqnarray} 
I_{2contr}^0=\frac{1}{2}\left(I_{a11}^0+I_{a12}^0\right)^2+I_{b11}^0\left(I_{a11}^0+I_{a12}^0\right)\approx 0.106
\end{eqnarray}
and
\begin{eqnarray} 
I_2^0=-1.4083307\pm 7.58*10^{-6},\quad I_2^1=-9.66805\pm 5.2*10^{-5},\quad I_3^0=-1.543\pm 0.0045 \nonumber \\
\end{eqnarray}
\begin{table}[t]
	\caption{\label{linearcontr}Linear in $b$ corrections to all two loop diagram contributions in the case of  $U(q)=\frac{1}{2}q^2+\frac{a}{3}q^3+\frac{b}{4}q^4,$ potential. Here, $s$ is the symmetry factor, $i$ - diagram label and $I_i$ - the diagram numerical value with absolute error ( $c11$ is calculated exactly).}
	\centering
	\begin{tabular}{|c|c|c|}
		\hline \multirow{1}{2cm}{~~~~~~~$i$}&\multirow{1}{3cm}{~~~~~~~$s$}&\multirow{1}{5cm}{$~~~~~~I_i^1$} \\
		\hline
		\multirow{1}{2cm}{~~~~a11}&\multirow{1}{3cm}{~~~~~$1/12$}&\multirow{1}{5cm}{$-0.748447\pm 1.5*10^{-5}$} \\
		\hline
		\multirow{1}{2cm}{~~~~a12}&\multirow{1}{3cm}{~~~~~$1/8$}&\multirow{1}{5cm}{$-0.981144\pm 4.8*10^{-5}$} \\
		\hline
		\multirow{1}{2cm}{~~~~b11}&\multirow{1}{3cm}{~~~~~$1/2$}&\multirow{1}{5cm}{$-8.338409\pm1.2*10^{-5}$} \\
		\hline
		\multirow{1}{2cm}{~~~~c11}&\multirow{1}{3cm}{~~~~~$1/8$}&\multirow{1}{5cm}{$6159/15400$} \\          
		\hline        
	\end{tabular}
\end{table}
\begin{table}[t]
	\caption{\label{AB12corrections2loop}Two loop contributions for the case $M=1, A=4, B=8$}
	\centering
	\begin{tabular}{|c|c|c|}
		\hline \multirow{1}{2cm}{~~~~~~~$i$}&\multirow{1}{3cm}{~~~~~~~~~~$s$}&\multirow{1}{5cm}{$~~~~~~I_i/S_b$} \\
		\hline 
		\multirow{1}{2cm}{~~~~~~~$a11$}&\multirow{1}{3cm}{~~~~~~~$~~1/12$}&\multirow{1}{5cm}{$-0.0023081\pm 1.1*10^{-6}$} \\
		\hline
		\multirow{1}{2cm}{~~~~~~~$a12$}&\multirow{1}{3cm}{~~~~~~~$~~1/8$}&\multirow{1}{5cm}{$-0.0310804\pm 2.5*10^{-6}$} \\
		\hline 
		\multirow{1}{2cm}{~~~~~~~$b11$}&\multirow{1}{3cm}{~~~~~~~$~~1/2$}&\multirow{1}{5cm}{$-0.0365297\pm 1.6*10^{-6}$} \\
		\hline  
		\multirow{1}{2cm}{~~~~~~~$c11$}&\multirow{1}{3cm}{~~~~~~~$~~1/8$}&\multirow{1}{5cm}{$0.00692251\pm 1.8*10^{-7}$} \\
		\hline    
	\end{tabular}
\end{table}
In the case $M=1, A=4, B=8$ the corresponding corrections are given by (see Tables \ref{AB12corrections2loop} and \ref{AB12corrections3loop} for the contributions of individual diagrams):
\begin{eqnarray} 
\text{Im}\,  E = \frac{2AB}{B-A}\frac{M^{3/4} A^{3/4} B^{3/4}}{\sqrt{\pi}}e^{-S_b}(1+\frac{I_2}{S_b}+\frac{I_3}{S_b^2})\, , 
\end{eqnarray}
where
\begin{eqnarray} 
I_2/S_b = (I_{a11}+I_{a12}+I_{b11}+I_{c11})/S_b = -0.0629957\pm 3.2*10^{-6},
\end{eqnarray}
and 
\begin{eqnarray} 
I_3/S_b^2=-0.0005401\pm 4.7*10^{-6},
\end{eqnarray}
We see that in the case with $M=1, A=4, B=8$ the higher order corrections are small and could be safety neglected. On the other hand in the limit $b\to 0$ the radiative corrections could be sizeable if $S_b\sim \mathcal{O}(1)$ and thus should be taken into account.
\begin{table}[t]
	\caption{\label{AB12corrections3loop} Three loop contributions for the case $M=1, A=4, B=8$. Here $I_{2loop}=\frac{1}{2}(I_{a11}+I_{a12}+I_{c11})^2+I_{b11}(I_{a11}+I_{a12}+I_{c11})$}
	\centering
	\begin{tabular}{|c|c|c|}
		\hline \multirow{1}{2cm}{~~~~~~~$i$}&\multirow{1}{3cm}{~~~~~~~$s$}&\multirow{1}{5cm}{$~~~~~~I_i/S_b^2$} \\
		\hline 
		\multirow{1}{2cm}{~~~~~~~$a21$}&\multirow{1}{3cm}{~~~~~~~$~~1/16$}
		&\multirow{1}{5cm}{$0.0000995\pm2.2*10^{-6} $} \\
		\hline
		\multirow{1}{2cm}{~~~~~~~$a22$}&\multirow{1}{3cm}{~~~~~~~$~~1/24$}
		&\multirow{1}{5cm}{$-0.00001402\pm2.3*10^{-7}  $} \\
		\hline 
		\multirow{1}{2cm}{~~~~~~~$a23$}&\multirow{1}{3cm}{~~~~~~~$~~1/16$}
		&\multirow{1}{5cm}{$0.00012921\pm6.1*10^{-7} $} \\
		\hline  
		\multirow{1}{2cm}{~~~~~~~$a24$}&\multirow{1}{3cm}{~~~~~~~$~~1/8$}
		&\multirow{1}{5cm}{$0.0000343\pm2.5*10^{-6}  $} \\
		\hline 
		\multirow{1}{2cm}{~~~~~~~$a25$}&\multirow{1}{3cm}{~~~~~~~$~~1/48$}
		&\multirow{1}{5cm}{$-0.0000647\pm1.1*10^{-6}  $} \\
		\hline
		\multirow{1}{2cm}{~~~~~~~$c21$}&\multirow{1}{3cm}{~~~~~~~$~~1/16$}
		&\multirow{1}{5cm}{$-0.000037825\pm9*10^{-9} $} \\
		\hline 
		\multirow{1}{2cm}{~~~~~~~$c22$}&\multirow{1}{3cm}{~~~~~~~$~~1/8$}
		&\multirow{1}{5cm}{$ 0.00010217\pm5.0*10^{-7} $} \\
		\hline  
		\multirow{1}{2cm}{~~~~~~~$c23$}&\multirow{1}{3cm}{~~~~~~~$~~1/8$}
		&\multirow{1}{5cm}{$-0.0002733\pm 3*10^{-7}  $} \\
		\hline  
		\multirow{1}{2cm}{~~~~~~~$c24$}&\multirow{1}{3cm}{~~~~~~~$~~1/12$}
		&\multirow{1}{5cm}{$-0.00031362\pm3.8*10^{-7}  $}\\
		\hline 
		\multirow{1}{2cm}{~~~~~~~$c25$}&\multirow{1}{3cm}{~~~~~~~$~~1/48$}
		&\multirow{1}{5cm}{$0.00018471\pm 1*10^{-8} $} \\
		\hline  
		\multirow{1}{2cm}{~~~~~~~$c26$}&\multirow{1}{3cm}{~~~~~~~$~~1/8$}
		&\multirow{1}{5cm}{$-0.00009389\pm2.0*10^{-7} $} \\
		\hline 
		\multirow{1}{2cm}{~~~~~~~$c27$}&\multirow{1}{3cm}{~~~~~~~$~~1/16$}
		&\multirow{1}{5cm}{$ 0.00020214\pm2.6*10^{-7} $}\\
		\hline 
		\multirow{1}{2cm}{~~~~~~~$b21$}&\multirow{1}{3cm}{~~~~~~~$~~1/4$}
		&\multirow{1}{5cm}{$-0.0007047\pm1.1*10^{-6}  $}\\
		\hline  
		\multirow{1}{2cm}{~~~~~~~$b22$}&\multirow{1}{3cm}{~~~~~~~$~~1/4$}
		&\multirow{1}{5cm}{$-0.0014675\pm1.6*10^{-6} $} \\
		\hline  
		\multirow{1}{2cm}{~~~~~~~$b23$}&\multirow{1}{3cm}{~~~~~~~$~~1/8$}
		&\multirow{1}{5cm}{$-0.0011163\pm2.2*10^{-6}  $} \\
		\hline  
		\multirow{1}{2cm}{~~~~~~~$d21$}&\multirow{1}{3cm}{~~~~~~~$~~1/4$}
		&\multirow{1}{5cm}{$0.0003165 \pm 4.2*10^{-7}$} \\
		\hline 
		\multirow{1}{2cm}{~~~~~~~$d22$}&\multirow{1}{3cm}{~~~~~~~$~~1/4$}
		&\multirow{1}{5cm}{$0.00086871\pm2.9*10^{-7}  $} \\
		\hline  
		\multirow{1}{2cm}{~~~~~~~$d23$}&\multirow{1}{3cm}{~~~~~~~$~~1/6$}
		&\multirow{1}{5cm}{$0.00029151\pm2.4*10^{-7}  $} \\
		\hline
		\multirow{1}{2cm}{~~$I_{2loop}/S_b^2$}&\multirow{1}{3cm}{~~~~~~~~$-$}
		&\multirow{1}{5cm}{$0.00131702\pm 1.1*10^{-7}$} \\
		\hline       
	\end{tabular}
\end{table}

\section{Conclusion}\label{ConclusionSec}

In this paper we considered the calculation of false vacuum decay rate for the case of arbitrary potential containing both cubic and quartic interactions in the framework of quantum mechanics. We obtained analytical expressions both for Green functions in the background of bounce solutions as well as for one-loop false vacuum decay rate. Besides, we provided numerical results for two and three-loop contributions to the decay rate together with a mathematica notebook, which could be used to evaluate two and three loop radiative corrections for arbitrary values of cubic and quartic couplings. The presented techniques could be further generalized to the case of quantum field theory, which will be our next step.   

\section*{Acknowledgements}

The authors would like to thank A.V. Bednyakov, A.E. Bolshov, M.Yu. Kalmykov,  D.I. Kazakov, D.I. Melikhov, V.N. Velizhanin, O.L. Veretin and A.F. Pikelner for interesting and stimulating discussions. This work was supported by RFBR grants \# 17-02-00872, \# 16-02-00943 and contract \# 02.A03.21.0003 from 27.08.2013 with Russian Ministry of Science and Education.


\begin{thebibliography}{99}


\bibitem{SMmetastability1} 
J.~Elias-Miro, J.~R.~Espinosa, G.~F.~Giudice, G.~Isidori, A.~Riotto and A.~Strumia,
{\it Higgs mass implications on the stability of the electroweak vacuum},
Phys.\ Lett.\ B {\bf 709}, 222 (2012)
[arXiv:1112.3022 [hep-ph]].

\bibitem{SMmetastability2} 
G.~Degrassi, S.~Di Vita, J.~Elias-Miro, J.~R.~Espinosa, G.~F.~Giudice, G.~Isidori and A.~Strumia,
{\it Higgs mass and vacuum stability in the Standard Model at NNLO},
JHEP {\bf 1208}, 098 (2012)
[arXiv:1205.6497 [hep-ph]].

\bibitem{SMmetastability3} 
F.~Bezrukov, M.~Y.~Kalmykov, B.~A.~Kniehl and M.~Shaposhnikov,
{\it Higgs Boson Mass and New Physics},
JHEP {\bf 1210}, 140 (2012)
[arXiv:1205.2893 [hep-ph]].


\bibitem{SMmetastability4} 
S.~Alekhin, A.~Djouadi and S.~Moch,
{\it The top quark and Higgs boson masses and the stability of the electroweak vacuum},
Phys.\ Lett.\ B {\bf 716}, 214 (2012)
[arXiv:1207.0980 [hep-ph]].


\bibitem{SMmetastability5} 
I.~Masina,
{\it Higgs boson and top quark masses as tests of electroweak vacuum stability},
Phys.\ Rev.\ D {\bf 87}, no. 5, 053001 (2013)
[arXiv:1209.0393 [hep-ph]].

\bibitem{SMmetastability6} 
D.~Buttazzo, G.~Degrassi, P.~P.~Giardino, G.~F.~Giudice, F.~Sala, A.~Salvio and A.~Strumia,
{\it Investigating the near-criticality of the Higgs boson},
JHEP {\bf 1312}, 089 (2013)
[arXiv:1307.3536 [hep-ph]].

\bibitem{SMmetastability7}
J.~R.~Espinosa, G.~F.~Giudice, E.~Morgante, A.~Riotto, L.~Senatore, A.~Strumia and N.~Tetradis,
{\it The cosmological Higgstory of the vacuum instability},
JHEP {\bf 1509}, 174 (2015)
[arXiv:1505.04825 [hep-ph]].

\bibitem{SMmetastability8} 
A.~V.~Bednyakov, B.~A.~Kniehl, A.~F.~Pikelner and O.~L.~Veretin,
{\it Stability of the Electroweak Vacuum: Gauge Independence and Advanced Precision},
Phys.\ Rev.\ Lett.\  {\bf 115}, no. 20, 201802 (2015)
[arXiv:1507.08833 [hep-ph]].

\bibitem{SMtunneling1}
G.~Isidori, G.~Ridolfi and A.~Strumia,
{\it On the metastability of the standard model vacuum},
Nucl.\ Phys.\ B {\bf 609}, 387 (2001)
[hep-ph/0104016].

\bibitem{SMtunneling2}
Z.~Lalak, M.~Lewicki and P.~Olszewski,
{\it Higher-order scalar interactions and SM vacuum stability},
JHEP {\bf 1405}, 119 (2014)
[arXiv:1402.3826 [hep-ph]].

\bibitem{SMtunneling3}
A.~D.~Plascencia and C.~Tamarit,
{\it Convexity, gauge-dependence and tunneling rates},
JHEP {\bf 1610}, 099 (2016)
[arXiv:1510.07613 [hep-ph]].

\bibitem{SMtunneling4}
M.~Endo, T.~Moroi, M.~M.~Nojiri and Y.~Shoji,
{\it Renormalization-Scale Uncertainty in the Decay Rate of False Vacuum},
JHEP {\bf 1601}, 031 (2016)
[arXiv:1511.04860 [hep-ph]].

\bibitem{SMtunneling5}
Z.~Lalak, M.~Lewicki and P.~Olszewski,
{\it Gauge fixing and renormalization scale independence of tunneling rate in Abelian Higgs model and in the standard model},
Phys.\ Rev.\ D {\bf 94}, no. 8, 085028 (2016)
[arXiv:1605.06713 [hep-ph]].

\bibitem{SMtunneling6}
O.~Czerwińska, Z.~Lalak, M.~Lewicki and P.~Olszewski,
{\it The impact of non-minimally coupled gravity on vacuum stability},
JHEP {\bf 1610}, 004 (2016)
[arXiv:1606.07808 [hep-ph]].

\bibitem{SMtunneling7}
J.~R.~Espinosa, M.~Garny, T.~Konstandin and A.~Riotto,
{\it Gauge-Independent Scales Related to the Standard Model Vacuum Instability},
Phys.\ Rev.\ D {\bf 95}, no. 5, 056004 (2017)
[arXiv:1608.06765 [hep-ph]].

\bibitem{SMtunneling8}
M.~Endo, T.~Moroi, M.~M.~Nojiri and Y.~Shoji,
{\it On the Gauge Invariance of the Decay Rate of False Vacuum},
Phys.\ Lett.\ B {\bf 771}, 281 (2017)
[arXiv:1703.09304 [hep-ph]].

\bibitem{SMtunneling9}
M.~Endo, T.~Moroi, M.~M.~Nojiri and Y.~Shoji,
{\it False Vacuum Decay in Gauge Theory},
arXiv:1704.03492 [hep-ph].


\bibitem{StochasticBackground1} 
E.~Witten,
{\it Cosmic Separation of Phases},
Phys.\ Rev.\ D {\bf 30}, 272 (1984).

\bibitem{StochasticBackground2} 
A.~Kosowsky, M.~S.~Turner and R.~Watkins,
{\it Gravitational radiation from colliding vacuum bubbles},
Phys.\ Rev.\ D {\bf 45}, 4514 (1992).

\bibitem{StochasticBackground3} 
A.~Kosowsky, M.~S.~Turner and R.~Watkins,
{\it Gravitational waves from first order cosmological phase transitions},
Phys.\ Rev.\ Lett.\  {\bf 69}, 2026 (1992).

\bibitem{StochasticBackground4} 
M.~Kamionkowski, A.~Kosowsky and M.~S.~Turner,
{\it Gravitational radiation from first order phase transitions},
Phys.\ Rev.\ D {\bf 49}, 2837 (1994)
[astro-ph/9310044].

\bibitem{StochasticBackground5} 
C.~Caprini, R.~Durrer, T.~Konstandin and G.~Servant,
Phys.\ Rev.\ D {\bf 79}, 083519 (2009)
doi:10.1103/PhysRevD.79.083519
[arXiv:0901.1661 [astro-ph.CO]].

\bibitem{GravitationalWavePhysics} 
R.~G.~Cai, Z.~Cao, Z.~K.~Guo, S.~J.~Wang and T.~Yang,
{\it The Gravitational Wave Physics},
arXiv:1703.00187 [gr-qc].


\bibitem{SakharovConditions} 
A.~D.~Sakharov,
{\it Violation of CP Invariance, c Asymmetry, and Baryon Asymmetry of the Universe},
Pisma Zh.\ Eksp.\ Teor.\ Fiz.\  {\bf 5}, 32 (1967)
[JETP Lett.\  {\bf 5}, 24 (1967)]
[Sov.\ Phys.\ Usp.\  {\bf 34}, 392 (1991)]
[Usp.\ Fiz.\ Nauk {\bf 161}, 61 (1991)].


\bibitem{BaryonAsymmetry1} 
D.~E.~Morrissey and M.~J.~Ramsey-Musolf,
{\it Electroweak baryogenesis},
New J.\ Phys.\  {\bf 14}, 125003 (2012)
[arXiv:1206.2942 [hep-ph]].

\bibitem{BaryonAsymmetry2} 
D.~J.~H.~Chung, A.~J.~Long and L.~T.~Wang,
{\it 125 GeV Higgs boson and electroweak phase transition model classes},
Phys.\ Rev.\ D {\bf 87}, no. 2, 023509 (2013)
[arXiv:1209.1819 [hep-ph]].

\bibitem{crossover1} 
K.~Kajantie, M.~Laine, K.~Rummukainen and M.~E.~Shaposhnikov,
{\it Is there a hot electroweak phase transition at m(H) larger or equal to m(W)?},
Phys.\ Rev.\ Lett.\  {\bf 77}, 2887 (1996)
[hep-ph/9605288].

\bibitem{crossover2} 
F.~Csikor, Z.~Fodor and J.~Heitger,
{\it Endpoint of the hot electroweak phase transition},
Phys.\ Rev.\ Lett.\  {\bf 82}, 21 (1999)
[hep-ph/9809291].

\bibitem{crossover3} 
Y.~Aoki, F.~Csikor, Z.~Fodor and A.~Ukawa,
{\it The Endpoint of the first order phase transition of the SU(2) gauge Higgs model on a four-dimensional isotropic lattice},
Phys.\ Rev.\ D {\bf 60}, 013001 (1999)
[hep-lat/9901021].

\bibitem{crossover4} 
M.~D'Onofrio and K.~Rummukainen,
{\it Standard model cross-over on the lattice},
Phys.\ Rev.\ D {\bf 93}, no. 2, 025003 (2016)
[arXiv:1508.07161 [hep-ph]].


\bibitem{PTcatalysis1} 
P.~Burda, R.~Gregory and I.~Moss,
{\it Gravity and the stability of the Higgs vacuum},
Phys.\ Rev.\ Lett.\  {\bf 115}, 071303 (2015)
[arXiv:1501.04937 [hep-th]].


\bibitem{PTcatalysis2} 
P.~Burda, R.~Gregory and I.~Moss,
{\it Vacuum metastability with black holes},
JHEP {\bf 1508}, 114 (2015)
[arXiv:1503.07331 [hep-th]].

\bibitem{PTcatalysis3} 
P.~Burda, R.~Gregory and I.~Moss,
{\it The fate of the Higgs vacuum},
JHEP {\bf 1606}, 025 (2016)
[arXiv:1601.02152 [hep-th]].

\bibitem{PTcatalysis4} 
D.~Gorbunov, D.~Levkov and A.~Panin,
{\it Fatal youth of the Universe: black hole threat for the electroweak vacuum during preheating},
arXiv:1704.05399 [astro-ph.CO].


\bibitem{Coleman1} 
S.~R.~Coleman,
{\it The Fate of the False Vacuum. 1. Semiclassical Theory},
Phys.\ Rev.\ D {\bf 15}, 2929 (1977)
Erratum: [Phys.\ Rev.\ D {\bf 16}, 1248 (1977)].

\bibitem{Coleman2} 
C.~G.~Callan, Jr. and S.~R.~Coleman,
{\it The Fate of the False Vacuum. 2. First Quantum Corrections},
Phys.\ Rev.\ D {\bf 16}, 1762 (1977).

\bibitem{Coleman3} 
S.~R.~Coleman and F.~De Luccia,
{\it Gravitational Effects on and of Vacuum Decay},
Phys.\ Rev.\ D {\bf 21}, 3305 (1980).

\bibitem{Kobzarev} 
I.~Y.~Kobzarev, L.~B.~Okun and M.~B.~Voloshin,
{\it Bubbles in Metastable Vacuum},
Sov.\ J.\ Nucl.\ Phys.\  {\bf 20}, 644 (1975)
[Yad.\ Fiz.\  {\bf 20}, 1229 (1974)].

\bibitem{KleinertPathIntegrals} 
H.~Kleinert,
{Path Integrals in Quantum Mechanics, Statistics, Polymer Physics, and Financial Markets},
World Scientific, Singapore, 2004

\bibitem{IntroQuantumMechanics} 
H.~J.~W.~Müller-Kirsten,
{\it Introduction to Quantum Mechanics : Schrödinger Equation and Path Integral},

\bibitem{QFTCriticalPhenomena} 
J.~Zinn-Justin,
{\it Quantum field theory and critical phenomena},
Int.\ Ser.\ Monogr.\ Phys.\  {\bf 113}, 1 (2002).

\bibitem{InstantonsLargeN} 
M.~Marino,
{\it Instantons and Large N : An Introduction to Non-Perturbative Methods in Quantum Field Theory},

\bibitem{WeinbergClassicalSolutions} 
E.~J.~Weinberg,
{\it Classical solutions in quantum field theory : Solitons and Instantons in High Energy Physics},

\bibitem{PrecisionDecayRates} 
A.~Andreassen, D.~Farhi, W.~Frost and M.~D.~Schwartz,
{\it Precision decay rate calculations in quantum field theory},
Phys.\ Rev.\ D {\bf 95}, no. 8, 085011 (2017)
[arXiv:1604.06090 [hep-th]].

\bibitem{DirectApproachQuantumTunneling} 
A.~Andreassen, D.~Farhi, W.~Frost and M.~D.~Schwartz,
{\it Direct Approach to Quantum Tunneling},
Phys.\ Rev.\ Lett.\  {\bf 117}, no. 23, 231601 (2016)
[arXiv:1602.01102 [hep-th]].

\bibitem{HeatKernelUserManual} 
D.~V.~Vassilevich,
{Heat kernel expansion: User's manual},
Phys.\ Rept.\  {\bf 388}, 279 (2003)
doi:10.1016/j.physrep.2003.09.002
[hep-th/0306138].

\bibitem{CalculationsExternalFields} 
V.~A.~Novikov, M.~A.~Shifman, A.~I.~Vainshtein and V.~I.~Zakharov,
{\it Calculations in External Fields in Quantum Chromodynamics. Technical Review},
Fortsch.\ Phys.\  {\bf 32}, 585 (1984).

\bibitem{MassiveContributionsQCDtunneling} 
O.~K.~Kwon, C.~k.~Lee and H.~Min,
{\it Massive field contributions to the QCD vacuum tunneling amplitude},
Phys.\ Rev.\ D {\bf 62}, 114022 (2000)
[hep-ph/0008028].


\bibitem{DunneFunctionalDeterminants} 
G.~V.~Dunne,
{\it Functional determinants in quantum field theory},
J.\ Phys.\ A {\bf 41}, 304006 (2008)
[arXiv:0711.1178 [hep-th]].


\bibitem{KleinertChervyakov1} 
H.~Kleinert and A.~Chervyakov,
{\it Functional determinants via Wronski construction of Green functions},
J.\ Math.\ Phys.\ {\bf 40}, 6044 (1999)
arXiv:physics/9712048 [math-ph].

\bibitem{KleinertChervyakov2} 
H.~Kleinert and A.~Chervyakov,
{\it Simple explicit formulas for Gaussian path integrals with time dependent frequencies},
Phys.\ Lett.\ A {\bf 245}, 345 (1998)
[quant-ph/9803016].

\bibitem{GarbrechtMillington1} 
B.~Garbrecht and P.~Millington,
{\it Green’s function method for handling radiative effects on false vacuum decay},
Phys.\ Rev.\ D {\bf 91}, 105021 (2015)
[arXiv:1501.07466 [hep-th]].

\bibitem{GarbrechtMillington2} 
B.~Garbrecht and P.~Millington,
{\it Self-consistent solitons for vacuum decay in radiatively generated potentials},
Phys.\ Rev.\ D {\bf 92}, 125022 (2015)
[arXiv:1509.08480 [hep-ph]].

\bibitem{GarbrechtMillington3} 
B.~Garbrecht and P.~Millington,
{\it Self-consistent radiative corrections to false vacuum decay},
arXiv:1703.05417 [hep-ph].


\bibitem{GenfaldYaglom} 
I.M.~Gel'fand and A.M.~Yaglom,
{\it Integration in Functional Spaces and its Applications in Quantum Physics},
J.\ Math.\ Phys.\ {\bf 1}, 48 (1960).

\bibitem{KirstenMcKane1} 
K.~Kirsten and A.~J.~McKane,
{\it Functional determinants by contour integration methods},
Annals Phys.\  {\bf 308}, 502 (2003)
[math-ph/0305010].


\bibitem{KirstenMcKane2} 
K.~Kirsten and A.~J.~McKane,
{\it Functional determinants for general Sturm-Liouville problems},
J.\ Phys.\ A {\bf 37}, 4649 (2004)
[math-ph/0403050].

\bibitem{ABCinstantons} 
A.~I.~Vainshtein, V.~I.~Zakharov, V.~A.~Novikov and M.~A.~Shifman,
{\it ABC's of Instantons},
Sov.\ Phys.\ Usp.\  {\bf 25}, 195 (1982)
[Usp.\ Fiz.\ Nauk {\bf 136}, 553 (1982)].

\bibitem{InstantonsQCD} 
T.~Schäfer and E.~V.~Shuryak,
{\it Instantons in QCD},
Rev.\ Mod.\ Phys.\  {\bf 70}, 323 (1998)
[hep-ph/9610451].

\bibitem{LecturesInstantons} 
S.~Vandoren and P.~van Nieuwenhuizen,
{\it Lectures on instantons},
arXiv:0802.1862 [hep-th].

\bibitem{AleinikovShuryak} 
A.~A.~Aleinikov and E.~V.~Shuryak,
{\it Instantons In Quantum Mechanics. Two Loop Effects. (in Russian)},
Yad.\ Fiz.\  {\bf 46}, 122 (1987).


\bibitem{Olejnik:1989id} 
S.~Olejnik,
{\it Do Nongaussian Effects Decrease Tunneling Probabilities? Three Loop Instanton Density For The Double Well Potential},
Phys.\ Lett.\ B {\bf 221}, 372 (1989).


\bibitem{Shuryak1} 
M.~A.~Escobar-Ruiz, E.~Shuryak and A.~V.~Turbiner,
{\it Three-loop Correction to the Instanton Density. I. The Quartic Double Well Potential},
Phys.\ Rev.\ D {\bf 92}, no. 2, 025046 (2015)
Erratum: [Phys.\ Rev.\ D {\bf 92}, no. 8, 089902 (2015)]
[arXiv:1501.03993 [hep-th]].

\bibitem{Shuryak2} 
M.~A.~Escobar-Ruiz, E.~Shuryak and A.~V.~Turbiner,
{\it Three-loop Correction to the Instanton Density. II. The Sine-Gordon potential},
Phys.\ Rev.\ D {\bf 92}, no. 2, 025047 (2015)
[arXiv:1505.05115 [hep-th]].

\bibitem{Shuryak3} 
M.~A.~Escobar-Ruiz, E.~Shuryak and A.~V.~Turbiner,
{\it Quantum and thermal fluctuations in quantum mechanics and field theories from a new version of semiclassical theory},
Phys.\ Rev.\ D {\bf 93}, no. 10, 105039 (2016)
[arXiv:1601.03964 [hep-th]].

\bibitem{EulerHeisenberg1} 
G.~V.~Dunne and C.~Schubert,
{\it Two loop selfdual Euler-Heisenberg Lagrangians. 1. Real part and helicity amplitudes},
JHEP {\bf 0208}, 053 (2002)
[hep-th/0205004].

\bibitem{EulerHeisenberg2} 
G.~V.~Dunne and C.~Schubert,
{\it Two loop selfdual Euler-Heisenberg Lagrangians. 2. Imaginary part and Borel analysis},
JHEP {\bf 0206}, 042 (2002)
[hep-th/0205005].

\bibitem{EulerHeisenberg3} 
G.~V.~Dunne,
{\it Heisenberg-Euler effective Lagrangians: Basics and extensions},
In *Shifman, M. (ed.) et al.: From fields to strings, vol. 1* 445-522
[hep-th/0406216].


\bibitem{Tseytlin1} 
R.~Roiban, A.~Tirziu and A.~A.~Tseytlin,
{\it Two-loop world-sheet corrections in AdS(5) x S**5 superstring},
JHEP {\bf 0707}, 056 (2007)
[arXiv:0704.3638 [hep-th]].

\bibitem{Tseytlin2} 
R.~Roiban and A.~A.~Tseytlin,
{\it Spinning superstrings at two loops: Strong-coupling corrections to dimensions of large-twist SYM operators},
Phys.\ Rev.\ D {\bf 77}, 066006 (2008)
[arXiv:0712.2479 [hep-th]].


\bibitem{CUBAlibrary} 
T.~Hahn,
{\it CUBA: A Library for multidimensional numerical integration},
Comput.\ Phys.\ Commun.\  {\bf 168}, 78 (2005)
[hep-ph/0404043].







\end{thebibliography}
\end{document}